\documentclass[pre,preprint,longbibliography]{revtex4-1}%
\usepackage{amssymb}
\usepackage{amsfonts}
\usepackage{amsmath}
\usepackage{xcolor}
\usepackage{ulem}
\usepackage{graphicx}%
\usepackage{bm}
\usepackage{dsfont}
\providecommand{\U}[1]{\protect\rule{.1in}{.1in}}
\newcommand{\tr}{\textcolor{red}}

\begin{document}
\title{Reconsidering Power Functional Theory}
\author{James F. Lutsko}
\homepage{http://www.lutsko.com}
\email{jlutsko@ulb.ac.be}
\affiliation{Center for Nonlinear Phenomena and Complex Systems CP 231, Universit\'{e}
Libre de Bruxelles, Blvd. du Triomphe, 1050 Brussels, Belgium}
\author{Martin Oettel}
\email{martin.oettel@uni-tuebingen.de}
\affiliation{Institute for Applied Physics, University of T{\"u}bingen, Auf der
Morgenstelle 10, 72076 T{\"u}bingen, Germany}

\begin{abstract}
  The original derivation of Power Functional Theory, Schmidt and Brader, JCP138,
  214101 (2013), is reworked in some detail with a view to clarifying and
simplifying the logic and making explicit the various functional dependencies. We note various issues
with the original development and suggest a modification that allows us to avoid them.
In the process we also suggest an alternative interpretation of our results
that bears surprising similarities to classical Density Functional Theory.
\end{abstract}
\date{\today }
\maketitle

\section{Introduction}

Classical Density\ Functional Theory (cDFT)\cite{Evans,lutsko} has proven to
be a powerful tool in the study of inhomogeneous classical systems. In his
seminal review article that helped to define the field, Evans also discussed a
dynamic extension that has come to be known as Dynamical Density Functional
Theory (DDFT) which is applicable to particles obeying an over-damped Brownian
dynamics.
The two are intimately related since the free energy functionals
that play a central role in cDFT are used in DDFT to describe the
deterministic driving force governing the evolution of the local
density\cite{Tarazona,Archer,Duran}. While cDFT is based on mathematically
exact theorems - and as such, represents a formally equivalent reformulation
of some aspects of classical statistical mechanics - DDFT is heuristic,
depending in all of its various derivations on a kind of uncontrolled,
local-equilibrium approximation. It would obviously be desirable if the
paradigm of DFT could be extended to non-equilibrium systems thus giving a
formally equivalent theory that, like cDFT, admits of simple but highly useful approximations.

This is precisely the goal of Power Functional Theory (PFT) as first proposed
by Schmidt and Brader\cite{SB1} (hereafter referred to as SB) for particles
subject to an over-damped Brownian dynamics and subsequently extended by
Schmidt and co-workers over the last several years to quantum\cite{SB3},
Newtonian\cite{SB2} and active-particle\cite{SB4} systems. The basic idea of
the various developments is to introduce a functional the minimization of
which generates the time-dependent many--particle distribution function for the system. A
version of constrained search (described below) is used to project this onto a
functional that depends only on the local density (as in cDFT) and the local
current, where the latter is the key to accessing time-dependent quantities. Minimization
with respect to these fields yields the actual time-dependent density and
current. A significant development inspired by this work, but not
directly dependent on it, has been the study of so-called
"non-adiabatic"\ dynamics - basically, the difference between some
non-equilibrium dynamics followed via simulation and the predictions of
DDFT\cite{S1,S2}. This has provided insight into the limits of DDFT\ (which
itself has become increasingly popular, see e.g. the recent review of te Vrugt,
L\"{o}wen and Wittkowski\cite{Lowen}) as well as some heuristic approaches to
modeling these differences. Recently, PFT has also been used to motivate
models of phase coexistence in active
particles\cite{Sactive1,Sactive2,Sactive3}.

In the following, we review the development of PFT using a notation that is
constructed to make particularly evident the functional dependencies which can
get lost with more standard, and less detailed, notation. 
The necessary calculations are presented in a fairly explicit manner to avoid ambiguities
as good as possible.
In doing so, certain
inconsistencies will be noted that, taken together, throw doubt on the
validity of the framework. {A modification aimed at overcoming these difficulties is proposed and yields new insights into the possibility of fulfilling  the program proposed by SB.} Section II of the paper sets the stage with a brief
review of Brownian dynamics and also introduces the functional notation.
Section III summarizes the key points of the exact PFT formalism including the
generating functional for the time-dependent $N$-body distribution function, its
projection onto the density-current subspace and some of the structural
elements stressed by SB and, in {the processes,} various problems in the derivation are
highlighted. In Section IV we present {our} modification of the original PFT of SB
that allows us to complete their program with surprising thoroughness. In the
process, we note an alternative interpretation of our results which may
provide a conceptual reformulation of Brownian Dynamics that is closer to the
goals of SB than was our original development. The paper ends with a brief
summary of our results.

\section{Brownian Dynamics}

The starting point is a system of $N$ identical, classical particles of mass
$m=1$ in $D$ dimensions for which the $i$-th particle has coordinates
$\mathbf{q}^{\left(  i\right)  }$ and velocities $\mathbf{v}^{\left(
i\right)  }$. The particles interact via a conservative potential
$U(\mathbf{q}^{(1)},...,\mathbf{q}^{(N)})$ which will be written more briefly
as $U(\mathbf{q}^{N})$ where $\mathbf{q}^{N}$ is the collection of all $N$
coordinates. In the following, a slightly compressed notation will be used
whereby the position (and time) arguments are written as subscripts so that
the potential will be written as $U_{\mathbf{q}^{N}}$. The particles also
experience a (possibly) time-dependent external one-body potential
$\phi_{t\mathbf{r}}$ . Again, what is written here as $\phi_{t\mathbf{r}}$
would, more conventionally be written as $\phi\left(  \mathbf{r},t\right)  $,
and in the following, the time argument will always precede the position
argument(s) in the subscript notation. Finally, there is a stochastic force
and corresponding friction proportional to the velocity (representing, e.g., a
bath of much smaller particles) so that the equations of motion are
\begin{align}
\frac{d}{dt}\widehat{\mathbf{q}}_{t}^{\left(  i\right)  } &  =\widehat
{\mathbf{v}}_{t}^{\left(  i\right)  }\\
\frac{d}{dt}\widehat{\mathbf{v}}_{t}^{\left(  i\right)  } &  =-\gamma
\widehat{\mathbf{v}}_{t}^{\left(  i\right)  }-\nabla_{i}U_{\widehat
{\mathbf{q}}_{t}^{N}}-\sum_{i=1}^{N}\nabla_{i}\phi_{t{\widehat{\mathbf{q}}%
_{t}^{\left(  i\right)  }}}+\widehat{\boldsymbol{\xi}}_{t}^{\left(  i\right)
}\nonumber
\end{align}
where a caret (hat) indicates a stochastic variable, $\gamma$ is the friction
and $\widehat{\boldsymbol{\xi}}_{t}^{\left(  i\right)  }$  are $D$-dimensional
vectors whose components are Gaussian-distributed white noise with
correlations $\left\langle \widehat{\boldsymbol{\xi}}_{t}^{\left(  i\right)
}\widehat{\boldsymbol{\xi}}_{t^{\prime}}^{\left(  j\right)  }\right\rangle
=2\gamma k_{B}T\delta_{ij}\delta\left(  t-t^{\prime}\right)  \mathds{1}$
(where $\mathds{1}$ is the unit tensor for the Cartesian components).
Following DDFT, the original PFT was developed for the over-damped limit in
which the time-derivative of the velocity can be neglected (this can be
justified rigorously with certain scaling assumptions, see e.g. Goddard, et
al.\cite{Goddard}) giving the so-called Brownian dynamics%
\begin{equation}
\frac{d}{dt}\widehat{\mathbf{q}}_{t}^{\left(  i\right)  }=\widehat{\mathbf{v}%
}_{t}^{\left(  i\right)  }=-\frac{1}{\gamma}\nabla_{i}U_{\widehat{\mathbf{q}%
}_{t}^{N}}-\frac{1}{\gamma}\sum_{i=1}^{N}\nabla_{i}\phi_{t{\widehat
{\mathbf{q}}_{t}^{\left(  i\right)  }}}+\frac{1}{\gamma}\widehat{\boldsymbol{\xi}%
}_{t}^{\left(  i\right)  }.
\end{equation}
The state of the system is entirely specified by the $N\times D$ phase space
coordinates $\widehat{\mathbf{q}}_{t}^{N}\equiv\widehat{\mathbf{q}}%
_{t}^{\left(  1\right)  },...,\widehat{\mathbf{q}}_{t}^{\left(  N\right)  }$.
In the following, it will be useful to note that the one-body term can be
written as
\begin{equation}
\sum_{i=1}^{N}\nabla_{i}\phi_{t{\widehat{\mathbf{q}}_{t}^{\left(  i\right)  }%
}}=\sum_{i=1}^{N}\nabla_{i}\left(  \sum_{j=1}^{N}\phi_{t{\widehat{\mathbf{q}%
}_{t}^{\left(  j\right)  }}}\right)  \equiv\left(  \sum_{i=1}^{N}\nabla
_{i}\right)  \phi_{t{\widehat{\mathbf{q}}_{t}^{N}}},\;
\end{equation}
where the last equivalence defines the total external potential,
$\phi_{t{\widehat{\mathbf{q}}_{t}^{N}}}$. The probability to find the system
in a given state, $\mathbf{r}^{N}$, at time $t$ is
\begin{equation}
\Psi_{t\mathbf{r}^{N}}=\left\langle \delta\left(  \mathbf{r}^{\left(
1\right)  }-\widehat{\mathbf{q}}_{t}^{\left(  1\right)  }\right)
...\delta\left(  \mathbf{r}^{\left(  N\right)  }-\widehat{\mathbf{q}}%
_{t}^{\left(  N\right)  }\right)  \right\rangle
\end{equation}
where the brackets indicate an average over the noise, $\widehat{\boldsymbol{\xi}%
}_{t}^{\left(  i\right)  }$, and the distribution of initial conditions. That
$\Psi_{t\mathbf{r}^{N}}$ is the distribution function is evident since the
expectation value with respect to the noise of any function $f_{\widehat
{\mathbf{q}}_{t}^{N}}$ of the stochastic variables $\widehat{\mathbf{q}}%
_{t}^{N}$ can be calculated as
\begin{equation}
\left\langle f_{\widehat{\mathbf{q}}_{t}^{N}}\right\rangle =\int
f_{\mathbf{r}^{N}}\Psi_{t\mathbf{r}^{N}}d\mathbf{r}^{N}%
\end{equation}
and as usual\cite{Gardiner}, the Brownian dynamics implies that the
distribution satisfies the Fokker-Planck equation%
\begin{equation}
\frac{\partial}{\partial t}\Psi_{t\mathbf{r}^{N}}=\frac{1}{\gamma}\sum
_{i=1}^{N}\nabla_{i} \cdot \left(  \nabla_{i}\left(  U_{\mathbf{r}^{N}}%
+\phi_{t\mathbf{r}^{N}}\right)  +k_{B}T\nabla_{i}\right)  \Psi_{t\mathbf{r}%
^{N}}.
\end{equation}
Clearly, the distribution at time $t$ is completely determined by the
interaction potential $U$, the external field $\phi_{t}$ and an initial
condition that will be denoted $\Psi_{t_{0}\mathbf{r}}$. As is to be expected, the
Boltzmann form $\Psi_{t\mathbf{r}^{N}}\sim\exp\left(  -\beta\left(
U_{\mathbf{r}^{N}}+\phi_{\mathbf{r}^{N}}\right)  \right)  $ is a stationary
solution if the field is stationary.

In the following, square brackets will be used to indicate functional
dependence so that to be precise and to indicate the full dependence of the
distribution on the interaction potential, the initial condition and the
external field one should write $\Psi_{t\mathbf{r}^{N}}\left[  U,\phi_{t}%
;\Psi_{0}\right]  $. Note that the spatial arguments are not indicated in the
square brackets because, in principle, to know $\Psi_{t\mathbf{r}^{N}}$ at
positions $\mathbf{r}^{N}$ requires knowing the various functional inputs
(e.g. $U,\phi_{t}$ and $\Psi_{0}$) at \emph{all} points in space. That said,
in the following the dependence on the inter-atomic potential $U$ will not be
explicitly indicated since it is ubiquitous, never changes and plays no role
in the discussion {below}. We note also that SB include non-conservative
forces in their analysis: although the same could be done here, we have chosen
to omit them for the sake of clarity as they do not change any of the
arguments to follow and only serve to complicate the expressions. We note for
later purposes that, in this notation, SB\ write the Fokker-Planck equation in
the form
\begin{equation}
\frac{\partial}{\partial t}\Psi_{t\mathbf{r}^{N}}=-\frac{1}{\gamma}\sum
	_{i=1}^{N}\nabla_{i} \cdot \left(  \mathbf{v}_{\mathbf{r}^{N}}^{\left(  i\right)
}\left[  \phi_{t},\Psi_{t}\right]  \Psi_{t\mathbf{r}^{N}}\right)  \label{FP}%
\end{equation}
with the "velocities" $\mathbf{v}_{\mathbf{r}^{N}}^{\left(  i\right)  }$ and
related "forces"\ $\mathbf{F}_{\mathbf{r}^{N}}^{\text{tot}\left(  i\right)  }$
defined as
\begin{equation}
\gamma\mathbf{v}_{\mathbf{r}^{N}}^{\left(  i\right)  }\left[  \phi
,\Psi\right]  =\mathbf{F}_{\mathbf{r}^{N}}^{\text{tot}\left(  i\right)
}\left[  \phi,\Psi\right]  =-\nabla_{i}U_{\mathbf{r}^{N}}-\nabla_{i}%
\phi_{\mathbf{r}^{N}}-k_{B}T\nabla_{i}\ln\Psi_{\mathbf{r}\tr{^{N}}}.
	\label{eq:velocities}
\end{equation}

The local number density (the probability to find a particle in a given
infinitesimal volume) is evaluated as 
\begin{equation}
\rho_{\mathbf{r}}\left[  \Psi\right]  = \sum_{i=1}^N \int\delta\left(  \mathbf{r}%
-\mathbf{r}^{\left(  i\right)  }\right)  \Psi_{\mathbf{r}^{N}}d\mathbf{r}^{N}%
\end{equation}
and, when the distribution - and so the density -  is time-dependent, the number current can be defined via the continuity equation,%
\begin{equation}
\frac{\partial}{\partial t}\rho_{\mathbf{r}}\left[  \Psi_{t}\right]
=-\nabla\cdot\mathbf{J}_{\mathbf{r}}\left[  \phi_{t},\Psi_{t}\right]
\end{equation}
which, together with the Fokker-Planck equation, gives the explicit
expression
\begin{equation}
\mathbf{J}_{\mathbf{r}}\left[  \phi,\Psi\right]  =\sum_{i=1}^{N}\int
\delta\left(  \mathbf{r}-\mathbf{r}^{\left(  i\right)  }\right)
\mathbf{v}_{\mathbf{r}^{N}}^{\left(  i\right)  }\left[  \phi,\Psi\right]
\Psi_{\mathbf{r}^{N}}d\mathbf{r}^{N}.\label{J}%
\end{equation}

An important distinction that is highlighted by this notation is that between
\emph{intrinsic} time dependence and \emph{inherited} time dependence.
Inherited time dependence refers to that arising simply because a functional
depends on another, time-dependent functional. For example, defining the
trivial functional $I_{\mathbf{r}}\left[  \phi\right]  =\phi_{\mathbf{r}}$,
its evaluation with a time-dependent input, e.g. $I_{\mathbf{r}}\left[
\phi_{t}\right]  $, has an inherited time-dependence coming solely from its
dependence on the time-dependent field. Here, the functional $I$ only changes
in time because its argument changes in time and otherwise, the time argument
is a passive label. An example here is the current $\mathbf{J}_{\mathbf{r}%
}\left[  \Psi,\phi\right]  $ for which the time dependence is inherited from
its arguments and which has no other source of time-dependence. On the other
hand, the distribution $\Psi_{t\mathbf{r}^{N}}\left[  \phi;\Psi_{0}\right]  $
has an \emph{intrinsic} time dependence since it is the solution to the
Fokker-Planck so that even if the external field is constant in time, $\Psi$
will still change in time if the initial condition is not the equilibrium
distribution. On the other hand, when the external field does change with
time, the distribution $\Psi_{t}$ depends on the values of the external field
$\phi_{\tau\mathbf{r}}$ at \emph{all} times $\tau<t$ and for this reason, one
writes $\Psi_{t\mathbf{r}^{N}}\left[  \phi;\Psi_{0}\right]  $ (with $\phi$
rather than $\phi_{t}$) because the functional acts on both the spatial
coordinate and the time.

This leads to a final caveat which is important throughout the analysis in\ SB
and below:\ namely, the role of causality. The Fokker-Planck equation can
formally be solved as
\begin{equation}
\Psi_{t\mathbf{r}^{N}}\left[  \phi;\Psi_{0}\right]  =\Psi_{t_{0}\mathbf{r}%
	^{N}}-\frac{1}{\gamma}\int_{t_{0}}^{t}\left\{  \sum_{i=1}^{N}\nabla_{i} \cdot \left(
\mathbf{v}_{\mathbf{r}^{N}}^{\left(  i\right)  }\left[  \phi_{t^{\prime}}%
,\Psi_{t^{\prime}}\right]  \Psi_{t^{\prime}\mathbf{r}^{N}}\right)  \right\}
dt^{\prime}%
\end{equation}
where $\Psi_{t_{0}\mathbf{r}^{N}}$ represents an initial value that has to be
specified. Notice that in the argument of $\Psi$ on the left hand side,  we write $\phi$ rather
than $\phi_{t}$ because the right hand side depends on the external field at
all times prior to $t$. SB specify that the integral on the right will be
understood to exclude the end-point at time $t$, so that $t_{0}\leq
t^{\prime}<t$ which can be interpreted as saying that the value of the field
at time $t$ is fully determined by its values at previous times corresponding,
physically, to the usual understanding of causality. When it is important below to note this dependence on a field at prior times, but not on the present time, we will write, e.g., $\Psi_{t\mathbf{r}^{N}}\left[  \phi_{<t};\Psi_{0}\right]$.

\section{Power Functional Theory}

\subsection{Variational formulation}

The analysis of SB begins with a a quantity modeled on the Rayleigh
dissipation function evaluated at a fixed time, $t$,
\begin{equation}
\widehat{R}_{\mathbf{r}^{N}}\left[  \widetilde{\mathbf{v}}^{N};\phi
_{t},\overset{\cdot}{\phi}_{t},\Psi_{t}\right]  =\sum_{i}\left(  \frac{\gamma
}{2}\widetilde{\mathbf{v}}_{\mathbf{r}^{N}}^{\left(  i\right)  }%
-\mathbf{F}_{\mathbf{r}^{N}}^{\text{tot}\left(  i\right)  }\left[  \phi
_{t},\Psi_{t}\right]  \right)  \cdot\widetilde{\mathbf{v}}_{\mathbf{r}^{N}%
}^{\left(  i\right)  }+\overset{\cdot}{\phi}_{t\mathbf{r}^{N}}%
\end{equation}
where $\widetilde{\mathbf{v}}_{\mathbf{r}^{N}}^{\left(  i\right)  }$ is a
collection of $N\ $test-fields (i.e. variational fields) each a function of
the $N$-positions $\mathbf{r}^{N}$. This is used to define the functional
\begin{equation}
R\left[  \widetilde{\mathbf{v}}^{N};\phi_{t},\overset{\cdot}{\phi}_{t}%
,\Psi_{t}\right]  =\int\Psi_{t\mathbf{r}^{N}}\widehat{R}_{\mathbf{r}^{N}%
}\left[  \widetilde{\mathbf{v}}^{N};\phi_{t},\overset{\cdot}{\phi}_{t}%
,\Psi_{t}\right]  d\mathbf{r}^{N}\label{R0}%
\end{equation}
which has the obvious property that its absolute, or global, minimum with
respect to the test fields $\widetilde{\mathbf{v}}^{N}$ occurs at
\begin{equation}
\gamma\widetilde{\mathbf{v}}_{\mathbf{r}^{N}}^{\left(  i\right)
\text{global-min}}\left[  \phi_{t},\Psi_{t}\right]  =\mathbf{F}_{\mathbf{r}%
^{N}}^{\text{tot}\left(  i\right)  }\left[  \phi_{t},\Psi_{t}\right]
=-\nabla_{i}U_{\mathbf{r}^{N}}-\nabla_{i}\phi_{t\mathbf{r}^{N}}-k_{B}%
T\nabla_{i}\ln\Psi_{t\mathbf{r}}\label{global}%
\end{equation}
corresponding to the "physical" fields at this fixed time $t$ that occur in
the Fokker-Planck equation as written in Eq.(\ref{FP}). SB then express the variational fields in terms of a new quantity, a variational distribution
$\widetilde{\Psi}_{\mathbf{r}^{N}}$ via the definition%
\begin{equation}
	\label{psitildedef}
\gamma\widetilde{\mathbf{v}}_{\mathbf{r}^{N}}^{\left(  i\right)  }%
\rightarrow\gamma\widetilde{\mathbf{v}}_{\mathbf{r}^{N}}^{\left(  i\right)
	}\left[ \phi,\widetilde{\Psi} \right]  =\mathbf{F}_{\mathbf{r}^{N}%
}^{\text{tot}\left(  i\right)  }\left[\phi,\widetilde{\Psi}\right]
\end{equation}
in terms of which the functional $R$ becomes (see Appendix \ref{appA})
\begin{equation}
R\left[  \widetilde{\Psi};\phi_{t},\overset{\cdot}{\phi}_{t},\Psi_{t}\right]
=\frac{\left(  k_{B}T\right)  ^{2}}{2\gamma}\int\sum_{i}\left(  \nabla_{i}%
\ln\frac{\widetilde{\Psi}_{\mathbf{r}^{N}}}{\Psi_{t\mathbf{r}^{N}}}\right)
^{2}\Psi_{t\mathbf{r}^{N}}d\mathbf{r}^{N}+\frac{1}{2}\frac{\partial}{\partial
t}\Lambda\left[  \phi_{t},\Psi_{t}\right]  +\frac{1}{2}\int\overset{\cdot
}{\phi}_{t\mathbf{r}^{N}}\Psi_{t\mathbf{r}^{N}}d\mathbf{r}^{N}\label{R}%
\end{equation}
where the dot-notation indicates a time-derivative and with%
\begin{equation}
\Lambda\left[  \phi,\Psi\right]  =\int\left(  k_{B}T\Psi_{\mathbf{r}^{N}}%
\ln\Psi_{\mathbf{r}^{N}}+\left(  U_{\mathbf{r}^{N}}+\phi_{\mathbf{r}^{N}%
}\right)  \Psi_{\mathbf{r}^{N}}\right)  d\mathbf{r}^{N}.
\end{equation}
Minimizing the generating function $R$ with respect to $\widetilde{\Psi}$
clearly gives $\widetilde{\Psi}^{\text{global-min}}=\Psi_{t}$, the physical
distribution. As observed by SB, the quantity $\Lambda$ plays a central role
in cDFT and is the same functional used by Mermin\cite{Mermin} to establish
its fundamental theorems. Here, it plays no role in the minimization procedure
and only serves to establish the value of generating function at its minimum.
Note that at this point one has not gained much yet:\ the variational velocity
field in Eq.(\ref{R0}) and the variational distribution in Eq.(\ref{R}) have
been introduced as extra fields, and the functional still depends on the
physical but unknown distribution $\Psi_{t\mathbf{r}^{N}}$.

\subsection{Power functional}

Next, SB break up the minimization of $R$ with respect to $\widetilde{\Psi}$
into two steps as
\begin{equation}
\min_{\widetilde{\Psi}}R\left[  \widetilde{\Psi};\phi_{t},\overset{\cdot}%
{\phi}_{t},\Psi_{t}\right]  =\min_{\overline{\rho},\overline{\mathbf{J}}%
}\left\{  \min_{\widetilde{\Psi}\rightarrow\overline{\rho},\overline
{\mathbf{J}}}R\left[  \widetilde{\Psi};\phi_{t},\overset{\cdot}{\phi}_{t}%
,\Psi_{t}\right]  \right\}  ,\label{split}%
\end{equation}
where the expression in curly brackets means that $R$ is first minimized with respect to the subset of possible
fields $\widetilde{\Psi}$ that satisfy $\rho_{\mathbf{r}}\left[
\widetilde{\Psi}\right]  =\overline{\rho}_{\mathbf{r}}$ and $\mathbf{J}%
_{\mathbf{r}}\left[  \phi_{t},\widetilde{\Psi}\right]  =\overline{\mathbf{J}%
}_{\mathbf{r}}$ for any given fields $\overline{\rho}_{\mathbf{r}}$ and $\overline{\mathbf{J}%
}_{\mathbf{r}}$. The first\ (inner) minimization defines a new functional%
\begin{equation}
\mathcal{R}\left[  \overline{\rho},\overline{\mathbf{J}};\phi_{t}%
,\overset{\cdot}{\phi}_{t},\Psi_{t}\right]  =\min_{\widetilde{\Psi}%
\rightarrow\overline{\rho},\overline{\mathbf{J}}}R\left[  \widetilde{\Psi
};\phi_{t},\overset{\cdot}{\phi}_{t},\Psi_{t}\right]  \label{d1}%
\end{equation}
and also defines one or more fields giving the minimum, $\widetilde{\Psi
}^{\text{min}}\left[  \overline{\rho},\overline{\mathbf{J}};\phi_{t}%
,\overset{\cdot}{\phi}_{t},\Psi_{t}\right]  ,$ so that
\begin{equation}
\mathcal{R}\left[  \overline{\rho},\overline{\mathbf{J}};\phi_{t}%
,\overset{\cdot}{\phi}_{t},\Psi_{t}\right]  =R\left[  \widetilde{\Psi
}^{\text{min}}\left[  \overline{\rho},\overline{\mathbf{J}};\phi_{t}%
,\overset{\cdot}{\phi}_{t},\Psi_{t}\right]  ;\phi_{t},\overset{\cdot}{\phi
}_{t},\Psi_{t}\right]
\end{equation}
In terms of the variational problem, it is important that $\overline{\rho
}_{\mathbf{r}}$ and $\overline{\mathbf{J}}_{\mathbf{r}}$ are \emph{arbitrary} fields
and that even though the current corresponding to $\widetilde{\Psi}$,  $\mathbf{J}_{\mathbf{r}}\left[  \phi_{t},\widetilde{\Psi
}\right]  $, carries an inherited time dependence via the external potential,
this is irrelevant to the choice of $\overline{\mathbf{J}}_{\mathbf{r}}$.
Instead, the time-dependence of the current manifests itself in that the
resulting functional $\mathcal{R}$ carries two dependencies on $\phi_{t}$: the
explicit dependence that comes via $\Lambda$ and, now, an \textit{implicit}
dependence that comes from minimizing under the constraint $\mathbf{J}%
_{\mathbf{r}}\left[  \phi_{t},\widetilde{\Psi}\right]  =\overline{\mathbf{J}%
}_{\mathbf{r}}$. For example, a toy model for $\mathcal{R}$, one that is
perfectly consistent and illustrates this difference, is
\begin{equation}
\mathcal{R}\left[  \overline{\rho},\overline{\mathbf{J}};\phi_{t}%
,\overset{\cdot}{\phi}_{t},\Psi_{t}\right]  \overset{?}{=}\int\left\{  \left(
\overline{\rho}_{\mathbf{r}}-\rho_{\mathbf{r}}\left[  \Psi_{t}\right]
\right)  ^{2}+\left(  \overline{\mathbf{J}}_{\mathbf{r}}-\mathbf{J}%
_{\mathbf{r}}\left[  \phi_{t},\Psi_{t}\right]  \right)  ^{2}\right\}
d\mathbf{r}+\frac{1}{2}\frac{\partial}{\partial t}\Lambda\left[  \phi_{t}%
,\Psi_{t}\right]  +\frac{1}{2}\int\overset{\cdot}{\phi}_{t\mathbf{r}^{N}}%
\Psi_{t\mathbf{r}^{N}}d\mathbf{r}^{N}\label{toy}%
\end{equation}
showing the explicit field dependence in $\Lambda$ as well as an implicit
field dependence in $\mathbf{J}_{\mathbf{r}}$. This toy-model minimizes (in the second step) to
give the correct, physical results $\overline{\rho}_{\mathbf{r}}%
=\rho_{\mathbf{r}}\left[  \Psi_{t}\right]  $ and $\overline{\mathbf{J}%
}_{\mathbf{r}}=\mathbf{J}_{\mathbf{r}}\left[  \phi_{t},\Psi_{t}\right]  $ and
this in turn (presumably)\ forces $\widetilde{\Psi}^{\text{global-min}}\left[
\rho\left[  \Psi_{t}\right]  ,\mathbf{J}_{\mathbf{r}}\left[  \phi_{t},\Psi
_{t}\right]  ;\Psi_{t},\phi_{t},\overset{\cdot}{\phi}_{t}\right]  =\Psi
_{t}\left[  \phi_{t},\overset{\cdot}{\phi}_{t}\right]  $. This happens with no
additional information as is the expected result of the two-step - or
"constrained search" - minimization procedure since this must ultimately lead to the same result
 as the one-step minimization, namely the exact
solution $\widetilde{\Psi}^{\text{global-min}}=\Psi_{t}$.

Following SB, the second step in the definition of the power functional is the
introduction of the quantity
\begin{align}
\overline{\Psi}_{t\mathbf{r}^{N}}\left[  \overline{\rho},\overline{\mathbf{J}%
};\phi,\Psi\right]   &  \equiv \Psi_{t_{0}\mathbf{r}^{N}}-\int_{t_{0}}%
^{t}\left\{  \sum_{i=1}^{N}\nabla_{i}\cdot\left(  \widetilde{\mathbf{v}%
}_{\mathbf{r}^{N}}^{\left(  i\right)  \text{min}}\left[  \phi_{t^{\prime}%
}\right]  \widetilde{\Psi}_{t^{\prime}\mathbf{r}^{N}}^{\text{min}}\left[
\overline{\rho}_{t^{\prime}},\overline{\mathbf{J}}_{t^{\prime}};\phi
_{t^{\prime}},\overset{\cdot}{\phi}_{t^{\prime}},\Psi_{t^{\prime}}\right]
\right)  \right\}  dt^{\prime}\label{c1}\\
&  =\Psi_{t_{0}\mathbf{r}^{N}}- \nonumber \\
	&\int_{t_{0}}^{t}\left\{  \sum_{i=1}^{N}%
\nabla_{i}\cdot\left(  \mathbf{v}_{\mathbf{r}^{N}}^{\left(  i\right)  }\left[
\widetilde{\Psi}^{\text{min}}\left[  \overline{\rho}_{t^{\prime}}%
,\overline{\mathbf{J}}_{t^{\prime}};\phi_{t^{\prime}},\overset{\cdot}{\phi
}_{t^{\prime}},\Psi_{t^{\prime}}\right]  ,\phi_{t^{\prime}}\right]
\widetilde{\Psi}_{t^{\prime}\mathbf{r}^{N}}^{\text{min}}\left[  \overline
{\rho},\overline{\mathbf{J}};\phi_{t^{\prime}},\overset{\cdot}{\phi
}_{t^{\prime}},\Psi_{t^{\prime}}\right]  \right)  \right\}  dt^{\prime
}\nonumber
\end{align}
which is a many--particle distribution fulfilling the continuity equation
if the velocities of the particles follow from the minimal trial distribution.
The first line is written as in SB, Eq.(18), whereas the second is in
our extended notation, showing that the minimum for the velocity fields
$\widetilde{\mathbf{v}}_{\mathbf{r}^{N}}^{\left(  i\right)  \text{min}}$
is determined by $\widetilde{\Psi}_{t^{\prime}\mathbf{r}^{N}}^{\text{min}}$
through Eq.~(\ref{psitildedef}). Notice first that a time-dependence has now been
assigned to the constraints $\overline{\rho}_{t^{\prime}},\overline
{\mathbf{J}}_{t^{\prime}}$. This means that they are being specified not at a
single moment but over some range of times $t_{0}\leq t\leq T$. Second, and
more problematic, notice that because $\widetilde{\Psi}_{t^{\prime}%
}^{\text{min}}$ is a result of Eq.(\ref{d1}), it has a dependence on the exact
distribution $\Psi_{t^{\prime}}$ evaluated at the same time $t^{\prime}$and so
$\overline{\Psi}_{t\mathbf{r}^{N}}$ {\it depends on the exact distribution at all
  earlier times}. While this appears to be the implication of equations (SB11),
(SB14) and (SB18)  it is at odds with subsequent developments in SB since the
point of this step is to eventually eliminate the exact distribution from this
problem. What seems to eventually be adopted is
the elimination of the exact solution $\Psi_t$ from the functional
through replacing
$\Psi_{t^{\prime}}$ by $\overline{\Psi}_{t^{\prime}}$, giving the final form of the
power functional,
\begin{align}
\overline{\mathcal{R}}\left[  \overline{\rho}_{t},\overline{\mathbf{J}}%
_{t};\phi_{t},\overset{\cdot}{\phi}_{t},\overline{\Psi}_{t}\right]   &
=\min_{\widetilde{\Psi}\rightarrow\overline{\rho}_{t},\overline{\mathbf{J}%
}_{t}}R\left[  \widetilde{\Psi};\phi_{t},\overset{\cdot}{\phi}_{t}%
,\overline{\Psi}_{t}\right]  \Rightarrow\widetilde{\Psi}^{\text{min}}\left[
\overline{\rho}_{t},\overline{\mathbf{J}}_{t};\phi_{t},\overset{\cdot}{\phi
}_{t},\overline{\Psi}_{t}\right]  \label{c2}\\
\overline{\Psi}_{t\mathbf{r}^{N}}\left[  \overline{\rho},\overline{\mathbf{J}%
};\phi,\overline{\Psi}\right]   &  =\Psi_{t_{0}\mathbf{r}^{N}}- \nonumber \\
	&\int_{t_{0}%
}^{t}\left\{  \sum_{i=1}^{N}\nabla_{i}\cdot\left(  \mathbf{v}_{\mathbf{r}^{N}%
}^{\left(  i\right)  }\left[  \widetilde{\Psi}^{\text{min}}\left[
\overline{\rho}_{t^{\prime}},\overline{\mathbf{J}}_{t^{\prime}};\phi
_{t^{\prime}},\overset{\cdot}{\phi}_{t^{\prime}},\overline{\Psi}_{t^{\prime}%
}\right]  ,\phi_{t^{\prime}}\right]  \widetilde{\Psi}_{t^{\prime}%
\mathbf{r}^{N}}^{\text{min}}\left[  \overline{\rho},\overline{\mathbf{J}}%
;\phi_{t^{\prime}},\overset{\cdot}{\phi}_{t^{\prime}},\overline{\Psi
}_{t^{\prime}}\right]  \right)  \right\}  dt^{\prime}\nonumber
\end{align}
The second line presents no difficulties in that evaluation of $\overline
{\Psi}_{t}$ requires $\widetilde{\Psi}^{\text{min}}$ at times $t^{\prime}<t$
and this in turn only requires $\overline{\Psi}_{t^{\prime}}$ at earlier
times. At this point, one can simplify the notation since everything at time
$t$ is now defined in terms of $\overline{\rho}$, $\overline{\mathbf{J}}$ at
all earlier times, so we will write a little more compactly
\begin{align}
\overline{\mathcal{R}}_{t}\left[  \overline{\rho},\overline{\mathbf{J}}%
;\phi_{t},\overset{\cdot}{\phi}_{t}\right]   &  =\min_{\widetilde{\Psi
}\rightarrow\overline{\rho}_{t},\overline{\mathbf{J}}_{t}}R\left[
\widetilde{\Psi};\phi_{t},\overset{\cdot}{\phi}_{t},\overline{\Psi}%
_{t}\right]  \Rightarrow\widetilde{\Psi}_{t}^{\text{min}}\left[
\overline{\rho},\overline{\mathbf{J}};\phi_{t},\overset{\cdot}{\phi}%
_{t}\right]  \label{c3}\\
\overline{\Psi}_{t\mathbf{r}^{N}}\left[  \overline{\rho},\overline{\mathbf{J}%
};\phi\right]   &  =\Psi_{t_{0}\mathbf{r}^{N}}-\int_{t_{0}}^{t}\left\{
\sum_{i=1}^{N}\nabla_{i}\cdot\left(  \mathbf{v}_{\mathbf{r}^{N}}^{\left(
i\right)  }\left[  \widetilde{\Psi}_{t^{\prime}}^{\text{min}}\left[
\overline{\rho},\overline{\mathbf{J}};\phi_{t^{\prime}},\overset{\cdot}{\phi
}_{t^{\prime}}\right]  ,\phi_{t^{\prime}}\right]  \widetilde{\Psi}_{t^{\prime
}\mathbf{r}^{N}}^{\text{min}}\left[  \overline{\rho},\overline{\mathbf{J}%
};\phi_{t^{\prime}},\overset{\cdot}{\phi}_{t^{\prime}}\right]  \right)
\right\}  dt^{\prime}\nonumber
\end{align}
The density implied by this $\overline{\Psi}$ is now easily evaluated from the
last of these with the result
\begin{equation}
\rho_{\mathbf{r}}\left[  \overline{\Psi}_{t}\right]  =\rho_{\mathbf{r}}\left[
  \Psi_{t_{0}}\right]  -\int_{t_{0}}^{t}\nabla\cdot \overline{\mathbf{J}%
}_{t^{\prime}\mathbf{r}}dt^{\prime}%
\end{equation}
which follows because, by definition, $\widetilde{\Psi}_{t^{\prime}%
}^{\text{min}}\left[  \overline{\rho},\overline{\mathbf{J}}\right]  $ is
precisely the distribution that implies the current $\overline{\mathbf{J}%
}_{t^{\prime}}$ while minimizing the generating function. So, the implied
density $\rho_{\mathbf{r}}\left[  \overline{\Psi}_{t}\right]  $ satisfies a
continuity equation with respect to the current $\overline{\mathbf{J}%
}_{t\mathbf{r}}$. One might have expected that $\rho_{\mathbf{r}}\left[
\overline{\Psi}_{t}\right]  $ should simply be $\overline{\rho}_{t\mathbf{r}}$
but this is clearly not the case since, at this point, the temporal evolution
of the fields  $\overline{\rho}_{t\mathbf{r}}$ and $\overline{\mathbf{J}%
}_{t\mathbf{r}}$ are completely independent. Indeed, one also sees that in
general $\mathbf{J}_{\mathbf{r}}\left[  \overline{\Psi}_{t};\phi_{t}\right]
\neq\overline{\mathbf{J}}_{t\mathbf{r}}$, i.e., the one--particle current from the
distribution $\bar\Psi$ is not the constraint current. This turns out to be important.

\subsection{Problems interpreting the power functional}

In SB, one of the main results is that the power functional has a Legendre
structure with, e.g., the density $\overline{\rho}_{t}$ and $\overset{\cdot
}{\phi}_{t}$ playing the role of conjugate variables thus mirroring the relation between the density and the external field in equilibrium DFT. In order to verify this,
we now turn to an examination of the power functional with the aim of making
the dependencies on the field at time $t$ explicit. 
Starting from the SB functional (Eq.~(\ref{R0})), performing the first minimization
with the replacement
$\Psi_{t\mathbf{r}^{N}} \to \overline{\Psi}_{t\mathbf{r}^{N}}$ (as described) and following the steps
detailed in Appendix \ref{appA} gives
\begin{align}
\overline{\mathcal{R}}_{t}\left[  \overline{\rho},\overline{\mathbf{J}}%
;\phi_{t},\overset{\cdot}{\phi}_{t}\right]   &  =\frac{\left(  k_{B}T\right)
^{2}\gamma}{2}\int\overline{\Psi}_{t\mathbf{r}^{N}}\left[  \overline{\rho
},\overline{\mathbf{J}};\phi\right]  \left(  \sum_{i}\left(  \nabla_{i}%
\ln\frac{\overline{\Psi}_{t\mathbf{r}^{N}}\left[  \overline{\rho}%
,\overline{\mathbf{J}};\phi\right]  }{\widetilde{\Psi}_{t\mathbf{r}^{N}%
}^{\text{min}}\left[  \overline{\rho},\overline{\mathbf{J}};\phi_{t}%
,\overset{\cdot}{\phi}_{t}\right]  }\right)  ^{2}\right)  d\mathbf{r}%
^{N}\label{W}\\
	&  -\frac{1}{2\gamma}\int\overline{\Psi}_{t\mathbf{r}^{N}}\left[
\overline{\rho},\overline{\mathbf{J}};\phi\right]  \sum_{i}\left(
\mathbf{F}_{\mathbf{r}^{N}}^{\text{tot}\left(  i\right)  }\left[
0,\overline{\Psi}_{t}\right]  \right)  ^{2}d\mathbf{r}^{N}\nonumber\\
&  -\frac{1}{2\gamma}\int\rho_{\mathbf{r}}\left[  \overline{\Psi}_{t}\right]
	\left(  \nabla\phi_{t\mathbf{r}}\right)  ^{2}d\mathbf{r}+%
\int\mathbf{J}_{\mathbf{r}}\left[  0,\overline{\Psi}_{t}\right]  \cdot\left(
	\nabla\phi_{\mathbf{r}}\right)  d\mathbf{r}+\int\overset{\cdot}{\phi
}_{t\mathbf{r}}\rho_{\mathbf{r}}\left[  \overline{\Psi}_{t}\right]
\;d\mathbf{r} .\nonumber
\end{align}
Here, all explicit dependence on the external potential at time $t$ has been
exposed.
SB argue that this can
be written as (see SB\ Eq.(25))
\begin{equation}
\overline{\mathcal{R}}_{t}^{\left(  SB\right)  }\left[  \overline{\rho
},\overline{\mathbf{J}};\phi,\overset{\cdot}{\phi}\right]  =W_{t}^{\left(
SB\right)  }\left[  \overline{\rho},\overline{\mathbf{J}};\phi,\overset{\cdot
	}{\phi}\right]  - \int\overline{\mathbf{J}}_{\mathbf{r}}%
	\cdot\left(  \nabla\phi_{t\mathbf{r}}\right)  d\mathbf{r}+\int\overset{\cdot
}{\phi}_{t\mathbf{r}}\overline{\rho}_{t\mathbf{r}}\;d\mathbf{r}%
\end{equation}
where $W_{t}^{\left(  SB\right)  }$ does not depend on the external field at
time $t$ but only at earlier times. This form is crucial later on in SB for identifying
a truly intrinsic functional of the dissipated power, similar to the intrinsic free energy
functional in equilibrium DFT.
Our form (\ref{W}) reveals three difficulties with that interpretation.
The first is that in Eq.~(\ref{W}) one
cannot guarantee that $\widetilde{\Psi}_{t\mathbf{r}^{N}}^{\text{min}}$ does
not depend on the external field at time $t$. Indeed, the global minimum of
$R$ certainly does (see Eq.~(\ref{global})) so it is entirely possible that the
constrained minimum $\widetilde{\Psi}_{t\mathbf{r}^{N}}^{\text{min}}$ does as
well. The second is that one sees in\ Eq.~(\ref{W}) an explicit, quadratic
dependence on the field at time $t$, which is ignored in SB . This seems to be
because SB switch between the representation of the variational field as
$\widetilde{\Psi}$ and the original formulation in terms of $\widetilde
{\mathbf{v}}^{N}$ and in terms of the latter there is no quadratic term.
However, formulating everything in terms of the velocities is not possible at
this point as the definition of $\overline{\Psi}_{t\mathbf{r}^{N}}$, see
Eqs.~(\ref{c2}) and (\ref{c3}),  involves both $\mathbf{v}_{\mathbf{r}^{N}%
}^{\left(  i\right)  }\left[  \widetilde{\Psi}_{t^{\prime}}^{\text{min}%
}...\right]  $, which could indeed be replaced by $\widetilde{\mathbf{v}}_{\mathbf{r}^{N}%
}^{\left(  i\right)  \text{min}}$ but also explicitly $\widetilde{\Psi
}_{t^{\prime}}^{\text{min}},$ which cannot. The third problem is that the term
involving the gradient of the field in the last line of Eq.(\ref{W}) is
written in terms of $\mathbf{J}_{\mathbf{r}}\left[  0,\overline{\Psi}%
_{t}\right]  $ and $\rho_{\mathbf{r}}\left[  \overline{\Psi}_{t}\right]  $
rather than $\mathbf{J}_{\mathbf{r}}\left[  \widetilde{\Psi}_{t}^{\text{min}%
};\phi_{t}\right]  =\overline{\mathbf{J}}_{t\mathbf{r}}$ and $\rho
_{\mathbf{r}}\left[  \widetilde{\Psi}_{t}^{\text{min}}\right]  =\overline
{\rho}_{t\mathbf{r}}$, as tacitly assumed in\ SB. This means, e.g., that SB
Eq.(26),
\begin{equation}
\frac{\delta}{\delta\overset{\cdot}{\phi}_{t\mathbf{r}}}\overline{\mathcal{R}%
}_{t}^{\left(  SB\right)  }\left[  \overline{\rho},\overline{\mathbf{J}}%
;\phi,\overset{\cdot}{\phi}\right]  =\overline{\rho}_{t\mathbf{r}},
\end{equation}
in which $\overline{\mathcal{R}}_{t}^{\left(  SB\right)  }$ is meant to act as
a generator for the density does not hold here. Finally, we observe that
several of these problems can be traced to the fact that  $\widetilde{\Psi
}_{t}^{\text{min}}$ and $\overline{\Psi}_{t}$ are not the same quantities - a
fact that we will exploit below to try to repair these problems. 

\subsection{Using the power functional}

Before exploring modifications of PFT, it is worthwhile to recall the final
purpose. SB eventually introduce an ansatz for the power functional which is
(SB Eq.(30))
\begin{equation}
\overline{\mathcal{R}}_{t}^{\left(  SB\right)  }\left[  \overline{\rho
},\overline{\mathbf{J}};\phi,\overset{\cdot}{\phi}\right]  =
	P_t[\overline{\rho},\overline{\mathbf{J}}] + 
\int \overline{\mathbf{J}}_{t\mathbf{r}}\cdot\nabla\frac{\delta
F\left[  \overline{\rho}_{t}\right]  }{\delta\overline{\rho}_{t\mathbf{r}}%
} d\mathbf{r}
	-\int\overline{\mathbf{J}}_{t\mathbf{r}%
	}\cdot\left(  \nabla\phi_{t\mathbf{r}}\right)  d\mathbf{r}+\int\overset{\cdot
}{\phi}_{t\mathbf{r}}\overline{\rho}_{t\mathbf{r}}\;d\mathbf{r.}%
	\label{eq:Pt}
\end{equation}
where $P_t[\overline{\rho},\overline{\mathbf{J}}]=W_{t}^{\left(
SB\right)  }\left[  \overline{\rho},\overline{\mathbf{J}}\right]
-\int \overline{\mathbf{J}}_{t\mathbf{r}}\cdot\nabla\frac{\delta
F\left[  \overline{\rho}_{t}\right]  }{\delta\overline{\rho}_{t\mathbf{r}}%
} d\mathbf{r} $ 
is the intrinsic functional of dissipated power.
This is then used to complete the minimization procedure defined in
Eq.(\ref{split}) giving the equations%
\begin{align}
\frac{\delta}{\delta\overline{\mathbf{J}}_{t\mathbf{r}}}\overline{\mathcal{R}%
}_{t}^{\left(  SB\right)  }\left[  \overline{\rho},\overline{\mathbf{J}}%
;\phi,\overset{\cdot}{\phi}\right]   &  =0\\
\frac{\delta}{\delta\overline{\rho}_{t\mathbf{r}}}\overline{\mathcal{R}}%
_{t}^{\left(  SB\right)  }\left[  \overline{\rho},\overline{\mathbf{J}}%
;\phi,\overset{\cdot}{\phi}\right]   &  =0\nonumber
\end{align}
however such an interpretation is untenable as it leads to unphysical results
(see Appendix \ref{appB}). In fact, SB say at this point that a Lagrange
multiplier should be introduced prior to minimization so as to enforce the
continuity equation relating the density and current. This statement is
problematic for three reasons. First, as illustrated above when discussing the
toy model,\ Eq.(\ref{toy}), the continuity equation is already implicit in the
formalism since it reproduces the exact distribution from which the continuity
equation for the density is automatically valid. Second, this ad-hoc
modification would not be the minimization that was defined in Eq.(\ref{split}%
) and that has been used throughout the analysis. Third, if $\overline{\rho
}_{t\mathbf{r}}$ and $\overline{\mathbf{J}}_{t\mathbf{r}}$ are related by the
continuity equation then they could never have been treated as independent
constraints - since the density is then fully determined by the temporal history of
the current -  and the original split of the minimization should have taken
the form
\begin{equation}
\min_{\widetilde{\Psi}}R\left[  \widetilde{\Psi};\phi_{t},\overset{\cdot}%
{\phi}_{t},\Psi_{t}\right]  =\min_{\overline{\mathbf{J}}}\left\{
\min_{\widetilde{\Psi}\rightarrow\overline{\mathbf{J}}}R\left[  \widetilde
{\Psi};\phi_{t},\overset{\cdot}{\phi}_{t},\Psi_{t}\right]  \right\}  ,
\end{equation}
with no density constraint at all and so leaving only the variational equation
with respect to the current. In such a case, with no variation with respect to
the density, the relationship with classical DFT  (as the equilibrium limit) becomes less
clear: it is relegated to the statement - based on the ansatz above - that the
current is zero, and so the density stationary in time, if the equilibrium
condition
\begin{equation}
\nabla\frac{\delta F\left[  \overline{\rho}_{t}\right]  }{\delta\overline
	{\rho}_{t\mathbf{r}}}=\nabla\mu=0
\end{equation}
for some constant $\mu$ holds.

\section{A variation on PFT}

The discussion above suggests that several of the problems identified can be
addressed with a few modifications of the theory, in particular the use of the
variational velocities rather than the variational distribution and imposing a
constraint on the current but not on the density. Since the distribution is
required in the definition of the current, some replacement for the
variational distribution $\widetilde\Psi$ must be found.
Recalling as one source of problems that $\mathbf{J}_{\mathbf{r}}\left[  \overline{\Psi}_{t};\phi_{t}\right]
\neq (\overline{\mathbf{J}}_{t\mathbf{r}}=
\mathbf{J}_{\mathbf{r}}\left[  \widetilde{\Psi}^\text{min}_{t};\phi_{t}\right])$, we begin
with a redefinition of $\overline{\Psi}_{t}$ as satisfying the equation
\begin{equation}
\frac{\partial}{\partial t}\overline{\Psi}_{t\mathbf{r}^{N}}\left[
	\widetilde{\mathbf{v}}^{N};\phi\right]  = \tr{-}\sum_{i}\nabla_{i}\cdot
\widetilde{\mathbf{v}}_{t\mathbf{r}^{N}}^{\left(  i\right)  }\overline{\Psi
}_{t\mathbf{r}^{N}}\left[  \widetilde{\mathbf{v}}^{N};\phi\right]
,\;\;\overline{\Psi}_{t_{0}\mathbf{r}^{N}}=\Psi_{t_{0}\mathbf{r}^{N}%
}\label{e1}%
\end{equation}
for any set of variational velocities $\widetilde{\mathbf{v}}_{t\mathbf{r}^{N}}^{\left(  i\right)  }$ specified for all relevant times.
This describes the time evolution of the many-body distribution due to an
arbitrary velocity field, the variational field $\widetilde{\mathbf{v}}^{N}$,
with initial condition  $\Psi_{t_{0}\mathbf{r}^{N}}$ that is the same initial
condition as for the exact distribution. This is of course equivalent to the
integral form
\begin{equation}
\overline{\Psi}_{t\mathbf{r}^{N}}\left[  \widetilde{\mathbf{v}}^{N}%
;\phi\right]  =\Psi_{t_{0}\mathbf{r}^{N}}-\int_{t_{0}}^{t}\left\{  \sum
_{i=1}^{N}\nabla_{i}\cdot\left(  \widetilde{\mathbf{v}}_{t^{\prime}%
\mathbf{r}^{N}}^{\left(  i\right)  }\overline{\Psi}_{t^{\prime}\mathbf{r}^{N}%
}\left[  \widetilde{\mathbf{v}}^{N};\phi\right]  \right)  \right\}
dt^{\prime}.
\end{equation}
We keep the original definition of $R$, Eq.(\ref{R0}), evaluated using
$\overline{\Psi}$ in place of the exact distribution giving
\begin{equation}
R\left[  \widetilde{\mathbf{v}}^{N};\phi_{t},\overset{\cdot}{\phi}%
_{t},\overline{\Psi}_{t}\right]  =\int_{t\mathbf{r}^{N}}\overline{\Psi
}_{t\mathbf{r}^{N}}\sum_{i}\left(  \frac{\gamma}{2}\widetilde{\mathbf{v}%
}_{\mathbf{r}^{N}}^{\left(  i\right)  }-\mathbf{F}_{\mathbf{r}^{N}%
}^{\text{tot}\left(  i\right)  }\left[  \phi_{t},\overline{\Psi}_{t}\right]
\right)  \cdot\widetilde{\mathbf{v}}_{\mathbf{r}^{N}}^{\left(  i\right)
}d\mathbf{r}^{N}+\int\overline{\Psi}_{t\mathbf{r}^{N}}\overset{\cdot}{\phi
}_{t\mathbf{r}^{N}}d\mathbf{r}^{N}%
\end{equation}
so that minimization will now give%
\begin{equation}
\widetilde{\mathbf{v}}_{\mathbf{r}^{N}}^{\left(  i\right)  \text{global-min}%
}\left[  \phi_{t},\overline{\Psi}_{t}\right]  =\frac{1}{\gamma}\mathbf{F}%
_{\mathbf{r}^{N}}^{\text{tot}\left(  i\right)  }\left[  \phi_{t}%
,\overline{\Psi}_{t}\right]  =-\nabla_{i}U_{\mathbf{r}^{N}}-\nabla_{i}%
\phi_{t\mathbf{r}^{N}}-k_{B}T\nabla_{i}\ln\overline{\Psi}_{t\mathbf{r}^{N}}.
\end{equation}
When this is substituted into the evolution equation, Eq.(\ref{e1}), it
becomes the Fokker-Planck equation and so, with the exact initial condition,
one recovers the exact distribution for the system thus demonstrating that
this is an exact reformulation of the problem.

The power functional is now defined using only a constraint on the current as
\begin{equation}
\mathcal{R}_{t}\left[  \overline{\mathbf{J}};\phi,\overset{\cdot}{\phi
}\right]  =\min_{\mathbf{J}\left[  \widetilde{\mathbf{v}}^{N};\overline{\Psi
}_{t}\right]  =\overline{\mathbf{J}}_{t}}R\left[  \widetilde{\mathbf{v}}%
^{N};\phi_{t},\overset{\cdot}{\phi}_{t},\overline{\Psi}_{t}^{\text{min}%
}\right]  =R\left[  \widetilde{\mathbf{v}}_{t}^{N\text{min}}\left[
	\overline{\mathbf{J}};\phi\right]  ;\phi_{t},\overset{\cdot}{\phi}%
_{t},\overline{\Psi}_{t}^{\text{min}}\left[ \overline{\mathbf{J}}_{<t} %
;\phi,\overset{\cdot}{\phi}\right]  \right]  \label{R1}%
\end{equation}
so here, the current is evaluated from the usual definition, Eq.(\ref{J}),
using $\widetilde{\mathbf{v}}^{N}$ and $\overline{\Psi}_{t}$ as inputs and its
value is constrained to be the specified $\overline{\mathbf{J}}_{t}$. At fixed
time $t$, the minimization is only a minimization with respect to
$\widetilde{\mathbf{v}}^{N}$ at time $t$ since $\overline{\Psi}_{t}%
^{\text{min}}$only depends on $\widetilde{\mathbf{v}}_{t^{\prime}%
}^{N\text{min}}$ at earlier times $t^{\prime}<t$. Also, we have written
$\overline{\Psi}_{t}^{\text{min}}\left[\overline{\mathbf{J}}_{<t}\right]  $
(equivalent to $\overline{\Psi}_{t}^{\text{min}}\left[  \widetilde{\mathbf{v}}^{N\text{min}}_{<t}\right]  $ )
because $\overline{\Psi}_{t}$ depends on $\widetilde{\mathbf{v}}_{t^{\prime
}\mathbf{r}^{N}}^{N}$ for earlier times and so we replace these with the
(independently determined) $\widetilde{\mathbf{v}}_{t^{\prime}\mathbf{r}^{N}%
}^{N\text{min}}\left[  \overline{\mathbf{J}}\right]  $ leaving only an
over-all dependence on the current temporal history. Analyzing this as before,
one finds the power functional
\begin{align}
\mathcal{R}_{t}\left[  \overline{\mathbf{J}}_{t};\phi_{t},\overset{\cdot}%
{\phi}_{t}\right]   &  =\int\overline{\Psi}_{t\mathbf{r}^{N}}^{\text{min}%
}\left[\overline{\mathbf{J}}_{<t};\phi,\overset{\cdot}{\phi}\right]  \left(
\sum_{i}\left(  \frac{\gamma}{2}\widetilde{\mathbf{v}}_{t\mathbf{r}^{N}%
}^{\left(  i\right)  \text{min}}\left[  \overline{\mathbf{J}},\phi
,\overset{\cdot}{\phi}\right]  -\mathbf{F}_{\mathbf{r}^{N}}^{\text{tot}\left(
i\right)  }\left[  0,\overline{\Psi}_{t}^{\text{min}}\right]  \right)
\cdot\widetilde{\mathbf{v}}_{t\mathbf{r}^{N}}^{\left(  i\right)  \text{min}%
}\left[  \overline{\mathbf{J}},\phi,\overset{\cdot}{\phi}\right]  \right)
d\mathbf{r}^{N}\\
	&  -\int\overline{\mathbf{J}}_{t\mathbf{r}}\cdot\left(  \nabla
	\phi_{t\mathbf{r}}\right)  d\mathbf{r}+\int\overset{\cdot}{\phi}_{t\mathbf{r}%
}\rho_{\mathbf{r}}\left[  \overline{\Psi}_{t}^{\text{min}}\right]
\;d\mathbf{r.}\nonumber
\end{align}
which has the form of SB\ Eq.~(25) consisting of the sum of a term independent
of the field plus linear dependencies on the field written in terms of the
current $\overline{\mathbf{J}}$ and the corresponding density. Thus, some of
the structural problems discussed above have been resolved, although once
again one cannot say at this point that $\widetilde{\mathbf{v}}_{t\mathbf{r}^{N}}^{\left(
i\right)  \text{min}}$ is independent of the field at time $t$.

One could continue by introducing an ansatz for the power functional as in SB,
but here we are able to do much more because it turns out that  - unlike in SB -  {\it the power
functional can be evaluated exactly}. Let us return to the basic definition in
Eq.(\ref{R1}) and note that the constrained minimization can be formulated
using a Lagrange parameter (really, a vector field $\boldsymbol{\lambda}_{\mathbf{r}}$) by first defining the
Lagrangian%
\begin{equation}
L\left[  \widetilde{\mathbf{v}}^{N};\phi_{t},\overset{\cdot}{\phi}%
_{t},\overline{\Psi}_{t}\right]  =R\left[  \widetilde{\mathbf{v}}^{N};\phi
_{t},\overset{\cdot}{\phi}_{t},\overline{\Psi}_{t}\right]  -\int
\boldsymbol{\lambda}_{\mathbf{r}}\cdot\left(  \overline{\mathbf{J}}_{t\mathbf{r}%
}-\mathbf{J}_{\mathbf{r}}\left[  \widetilde{\mathbf{v}}^{N};\phi_{t},\overline{\Psi
}_{t}\right]  \right)  d\mathbf{r}%
\end{equation}
and then minimizing by solving
\begin{align}
0 &  =\frac{\delta}{\delta\widetilde{\mathbf{v}}_{\mathbf{r}}^{\left(
i\right)  }}L\left[  \widetilde{\mathbf{v}}^{N};\phi_{t},\overset{\cdot}{\phi
}_{t},\overline{\Psi}_{t}^{\text{min}}\right]  \\
	0 &  =\frac{\delta}{\delta\boldsymbol{\lambda}_{\mathbf{r}}}L\left[  \widetilde{\mathbf{v}}^{N}%
;\phi_{t},\overset{\cdot}{\phi}_{t},\overline{\Psi}_{t}^{\text{min}}\right]
\nonumber
\end{align}
which gives, since $R$ is quadratic in the velocities,%
\begin{align}
0 &  =\overline{\Psi}_{t\mathbf{r}^{N}}\left[  \widetilde{\mathbf{v}}_{t}%
^{N},\phi\right]  \left(  \gamma\widetilde{\mathbf{v}}_{t\mathbf{r}^{N}%
}^{\left(  i\right)  }-\mathbf{F}_{\mathbf{r}^{N}}^{\text{tot}\left(
i\right)  }\left[  \phi_{t},\overline{\Psi}_{t}^{\text{min}}\right]  \right)
+\int\left(  \frac{\delta}{\delta\widetilde{\mathbf{v}}_{\mathbf{r}}^{\left(
i\right)  }}\mathbf{J}_{\mathbf{r}}\left[  \widetilde{\mathbf{v}}^{N};\phi
_{t},\overline{\Psi}_{t}^{\text{min}}\right]  \right)  \cdot\boldsymbol{\lambda
}_{\mathbf{r}}d\mathbf{r}\\
0 &  =\overline{\mathbf{J}}_{t\mathbf{r}}-\mathbf{J}_{\mathbf{r}}\left[
\widetilde{\mathbf{v}}^{N};\phi_{t},\overline{\Psi}_{t}^{\text{min}}\right]
\nonumber
\end{align}
Now, it is straightforward to evaluate
\begin{equation}
\int\left(  \frac{\delta}{\delta\widetilde{\mathbf{v}}_{\mathbf{r}}^{\left(
i\right)  }}\mathbf{J}_{\mathbf{r}}\left[  \widetilde{\mathbf{v}}^{N};\phi
_{t},\overline{\Psi}_{t}^{\text{min}}\right]  \right)  \cdot\boldsymbol{\lambda
}_{\mathbf{r}}d\mathbf{r=}\int\boldsymbol{\lambda}_{\mathbf{r}}\delta\left(
\mathbf{r}-\mathbf{r}_{i}\right)  \overline{\Psi}_{t\mathbf{r}^{N}}^{\text{min}}d\mathbf{r}
	=\boldsymbol{\lambda}_{\mathbf{r}_{i}}\overline{\Psi}_{t\mathbf{r}^{N}}^{\text{min}}%
\end{equation}
so that the system of equations becomes%
\begin{align}
0 &  =\overline{\Psi}_{t\mathbf{r}^{N}}^{\text{min}}\left[  \widetilde
{\mathbf{v}}_{t}^{N},\phi\right]  \left(  \gamma\widetilde{\mathbf{v}%
}_{t\mathbf{r}^{N}}^{\left(  i\right)  }-\mathbf{F}_{\mathbf{r}^{N}%
}^{\text{tot}\left(  i\right)  }\left[  \phi_{t},\overline{\Psi}%
_{t}^{\text{min}}\right]  \right)  +\boldsymbol{\lambda}_{\mathbf{r}_{i}}%
\overline{\Psi}_{t\mathbf{r}^{N}}^{\text{min}}\left[  \widetilde{v}_{t}%
^{N},\phi\right]  \\
0 &  =\overline{\mathbf{J}}_{t\mathbf{r}}-\mathbf{J}_{\mathbf{r}}\left[
\widetilde{\mathbf{v}}^{N};\phi_{t},\overline{\Psi}_{t}^{\text{min}}\right]
.\nonumber
\end{align}
The first line gives
\begin{equation}
\widetilde{\mathbf{v}}_{t\mathbf{r}^{N}}^{\left(  i\right)  \text{min}}\left[
\overline{\mathbf{J}},\phi,\overset{\cdot}{\phi}\right]  =\frac{1}{\gamma
}\mathbf{F}_{\mathbf{r}^{N}}^{\text{tot}\left(  i\right)  }\left[  \phi
_{t},\overline{\Psi}_{t}^{\text{min}}\right]  -\frac{1}{\gamma}\boldsymbol{\lambda
}_{\mathbf{r}_{i}}%
\end{equation}
while the second then evaluates to
\begin{equation}
\overline{\mathbf{J}}_{t\mathbf{r}}=\mathbf{J}_{\mathbf{r}}\left[
\widetilde{\mathbf{v}}^{N\text{min}};\phi_{t},\overline{\Psi}_{t}^{\text{min}%
}\right]  =\mathbf{J}_{\mathbf{r}}\left[  \phi_{t},\overline{\Psi}%
_{t}^{\text{min}}\right]  -\frac{1}{\gamma}\boldsymbol{\lambda}_{\mathbf{r}}%
\rho_{\mathbf{r}}\left[  \overline{\Psi}_{t}^{\text{min}}\right]
\end{equation}
so that the solution to the minimization problem is
\begin{align}
\boldsymbol{\lambda}_{\mathbf{r}}^{\text{min}} &  =\gamma\frac{\mathbf{J}%
_{\mathbf{r}}\left[  \phi_{t},\overline{\Psi}_{t}^{\text{min}}\right]
-\overline{\mathbf{J}}_{t\mathbf{r}}}{\rho_{\mathbf{r}}\left[  \overline{\Psi
}_{t}^{\text{min}}\right]  }\label{min}\\
\widetilde{\mathbf{v}}_{t\mathbf{r}^{N}}^{\left(  i\right)  \text{min}}\left[
\overline{\mathbf{J}},\phi,\overset{\cdot}{\phi}\right]   &  =\frac{1}{\gamma
}\mathbf{F}_{\mathbf{r}^{N}}^{\text{tot}\left(  i\right)  }\left[  \phi
_{t},\overline{\Psi}_{t}^{\text{min}}\right]  -\frac{\mathbf{J}_{\mathbf{r}%
_{i}}\left[  \phi_{t},\overline{\Psi}_{t}^{\text{min}}\right]  -\overline
{\mathbf{J}}_{t\mathbf{r}_{i}}}{\rho_{\mathbf{r}_{i}}\left[  \overline{\Psi
}_{t}^{\text{min}}\right]  }.\nonumber
\end{align}
Using this to evaluate the power functional gives
\begin{eqnarray}
\mathcal{R}_{t}\left[  \overline{\mathbf{J}}_{t};\phi_{t},\overset{\cdot}%
	{\phi}_{t}\right] & =&-\frac{1}{2\gamma}\int\overline{\Psi}_{t\mathbf{r}^{N}%
}^{\text{min}}\sum_{i}\left(  \mathbf{F}_{\mathbf{r}^{N}}^{\text{tot}\left(
i\right)  }\left[  \phi_{t},\overline{\Psi}_{t}^{\text{min}}\right]  \right)
^{2}d\mathbf{r}^{N}+\frac{\gamma}{2}\int\frac{\left(  \overline{\mathbf{J}%
}_{t\mathbf{r}}-\mathbf{J}_{\mathbf{r}}\left[  \phi_{t},\overline{\Psi}%
_{t}^{\text{min}}\right]  \right)  ^{2}}{\rho_{\mathbf{r}}\left[
\overline{\Psi}_{t}^{\text{min}}\right]  }d\mathbf{r} \nonumber\\
 & &+\int\overset{\cdot}%
{\phi}_{t\mathbf{r}}\rho_{\mathbf{r}}\left[  \overline{\Psi}_{t}^{\text{min}%
}\right]  \;d\mathbf{r}.%
\end{eqnarray}
The dependence on the external field $\phi_t$ appearing in $\mathbf{F}_{\mathbf{r}^{N}}^{\text{tot}\left(
i\right)  }\left[  \phi_{t},\overline{\Psi}_{t}^{\text{min}}\right]$ and
$\mathbf{J}_{\mathbf{r}}\left[  \phi_{t},\overline{\Psi}_{t}^{\text{min}}\right]$
can be explicitly taken out and the functional is rewritten as
\begin{eqnarray}
\mathcal{R}_{t}\left[  \overline{\mathbf{J}}_{t};\phi_{t},\overset{\cdot}%
	{\phi}_{t}\right]  &=&-\frac{1}{2\gamma}\int\overline{\Psi}_{t\mathbf{r}^{N}%
}^{\text{min}}\sum_{i}\left(  \mathbf{F}_{\mathbf{r}^{N}}^{\text{tot}\left(
i\right)  }\left[  0,\overline{\Psi}_{t}^{\text{min}}\right]  \right)
^{2}d\mathbf{r}^{N}+\frac{\gamma}{2}\int\frac{\left(  \overline{\mathbf{J}%
}_{t\mathbf{r}}-\mathbf{J}_{\mathbf{r}}\left[  0,\overline{\Psi}%
_{t}^{\text{min}}\right]  \right)  ^{2}}{\rho_{\mathbf{r}}\left[
\overline{\Psi}_{t}^{\text{min}}\right]  }d\mathbf{r} \nonumber \\
	 & & -\int
\overline{\mathbf{J}}_{t\mathbf{r}}\cdot\left(  \nabla\phi_{t\mathbf{r}%
	}\right)  d\mathbf{r}+\int\overset{\cdot}{\phi}_{t\mathbf{r}}\rho_{\mathbf{r}%
}\left[  \overline{\Psi}_{t}^{\text{min}}\right]  \;d\mathbf{r,}%
\end{eqnarray}
thus explicitly showing the structure discussed in\ SB whereby the external
field contributions at time $t$ appear linearly. Finally, we note that the
density automatically satisfies the continuity equation since
\begin{align}
\frac{\partial}{\partial t}\rho_{\mathbf{r}}\left[  \overline{\Psi}%
_{t}^{\text{min}}\right]   &  =\int\frac{\partial}{\partial t}\overline{\Psi
}_{t\mathbf{r}^{N}}^{\text{min}}\sum_{i}\delta\left(  \mathbf{r}%
-\mathbf{r}_{i}\right)  d\mathbf{r}^{N}\\
	&  =-\int\sum_{i}\delta\left(  \mathbf{r}-\mathbf{r}_{i}\right)  \nabla
_{i}\cdot\left(  \widetilde{\mathbf{v}}_{t\mathbf{r}^{N}}^{\left(  i\right)
\text{min}}\overline{\Psi}_{t\mathbf{r}^{N}}^{\text{min}}\right)
d\mathbf{r}^{N}\nonumber\\
&  =-\nabla\cdot\int\sum_{i}\delta\left(  \mathbf{r}-\mathbf{r}_{i}\right)
\widetilde{\mathbf{v}}_{t\mathbf{r}^{N}}^{\left(  i\right)  \text{min}%
}\overline{\Psi}_{t\mathbf{r}^{N}}^{\text{min}}d\mathbf{r}^{N}\nonumber\\
&  =-\nabla\cdot\overline{\mathbf{J}}_{t\mathbf{r}}\nonumber
\end{align}
by definition of $\widetilde{\mathbf{v}}_{t\mathbf{r}^{N}}^{\left(  i\right)
\text{min}}$. This shows that we can replace $\rho_{\mathbf{r}}\left[
\overline{\Psi}_{t}^{\text{min}}\right]  $ by $\rho_{t\mathbf{r}}\left[
\overline{\mathbf{J}}\right]  $ since the current (and an initial condition)
completely determines the density. We can interpret this construction as
follows: $\mathcal{R}_{t}\left[  \overline{\mathbf{J}}_{t};\phi_{t}%
,\overset{\cdot}{\phi}_{t}\right]  $ is a functional for the power in a system
which is described by a certain time-dependent current $\overline{\mathbf{J}%
}_{t}$. The associated many-body distribution $\overline{\Psi}_{t}%
^{\text{min}}\left[ \overline{\mathbf{J}}_{<t} \right]  $ is consistent with this
current (i.e. they are related by the usual definition) and the corresponding
local density fulfills  the continuity with  $\overline{\mathbf{J}}_{t}$ as
the material-transporting current. Upon minimization with respect to
$\overline{\mathbf{J}}_{t}$, under the usual constraint of causality, the
minimal current $\overline{\mathbf{J}}_{t}^{\text{min }}$and $\overline{\Psi
}_{t}^{\text{min}}\left[ \overline{\mathbf{J}}_{<t} \right]  $ become
the exact physical solutions corresponding to the applied field $\phi_{t}$. 

\subsection{Pessimistic interpretation of these results}

Following SB, the power functional is to be minimized with respect to the
current so as to get the minimizing current. Here, this is trivial and results
in
\begin{equation}
	\label{eq:min1}
\overline{\mathbf{J}}_{t\mathbf{r}}^{\text{min}}=\mathbf{J}_{\mathbf{r}%
	}\left[  \phi_{t},\overline{\Psi}_{t}^{\text{min}}[\overline{\mathbf{J}}_{<t}]\right]
\end{equation}
and from\ Eq.(\ref{min}), one then has for the variational velocities%
\begin{equation}
\widetilde{\mathbf{v}}_{t\mathbf{r}^{N}}^{\left(  i\right)  \text{min}}\left[
\overline{\mathbf{J}}_{t\mathbf{r}}^{\text{min}},\phi,\overset{\cdot}{\phi
}\right]  =\frac{1}{\gamma}\mathbf{F}_{\mathbf{r}^{N}}^{\text{tot}\left(
i\right)  }\left[  \phi_{t},\overline{\Psi}_{t}^{\text{min}}\right]  .
\end{equation}
As demonstrated at the start of this section (see Eq.~(\ref{e1})), this simply implies that
$\overline{\Psi}_{t}^{\text{min}}$ satisfies the original Fokker-Planck
equation and, so, is the exact distribution. 
One arrives at the same conclusion by recognizing that the solution to Eq.~(\ref{eq:min1}) for the current at
time $t$ depends on the current at all earlier times, and for finding a self--consistent solution
($\overline{\mathbf{J}}_{<t} = \overline{\mathbf{J}}_{<t}^\text{min}$) one has to solve Eq.~(\ref{eq:min1}) at all
earlier times successively. This amounts to nothing but the solution of the original Fokker-Planck
equation.

Without the freedom to introduce
an ansatz by hand, the formalism just reduces to the original exact result.
There seems, therefore, to be no advantage to this development since, while
one could introduce one of the usual approximations (e.g. local equilibrium)
at any point, there seems no particular rationale to do so as opposed to,
e.g., simply inserting such an ansatz directly into the continuity equation
for the density (as is done in heuristic derivations of DDFT\cite{Lowen}).

\subsection{An optimistic interpretation}

A somewhat different interpretation of the formalism is possible upon
consideration of the evolution equation for $\overline{\Psi}_{t}^{\text{min}}%
$, evaluated with the velocities constrained by the current,\ Eq.(\ref{min}),
\begin{eqnarray}
\frac{\partial}{\partial t}\overline{\Psi}_{t\mathbf{r}^{N}}^{\text{min}%
	}\left[\overline{\mathbf{J}}_{<t},\phi\right]  &=& - \sum_{i}\nabla_{i}\cdot\frac
{1}{\gamma}\mathbf{F}_{\mathbf{r}^{N}}^{\text{tot}\left(  i\right)  }\left[
\phi_{t},\overline{\Psi}_{t}^{\text{min}}\right]  \overline{\Psi}%
_{t\mathbf{r}^{N}}^{\text{min}}\left[  
	\overline{\mathbf{J}}_{<t}
,\phi\right]  \nonumber \\
	&& + \sum_{i}\nabla_{i}\cdot\left(  \frac{\mathbf{J}_{\mathbf{r}%
_{i}}\left[  \overline{\Psi}_{t}^{\text{min}},\phi_{t}\right]  -\overline
{\mathbf{J}}_{t\mathbf{r}_{i}}}{\rho_{t\mathbf{r}_{i}}\left[  \overline
{\mathbf{J}}\right]  }\; \overline{\Psi}_{t\mathbf{r}^{N}}^{\text{min}}\left[
	\overline{\mathbf{J}}_{<t},\phi\right]  \right)  ,\;\;\overline{\Psi}_{t_{0}\mathbf{r}^{N}%
}^\text{{min}}=\Psi_{t_{0}\mathbf{r}^{N}}%
\end{eqnarray}
and it is straightforward to show that the {explicit} contribution of the external field
{from $\mathbf{F}_{\mathbf{r}^{N}}^{\text{tot}}$} cancels on the right hand side leaving
\begin{eqnarray}
\frac{\partial}{\partial t}\overline{\Psi}_{t\mathbf{r}^{N}}^{\text{min}%
	}\left[{\overline{\mathbf{J}}_{<t}},\phi\right]  &=&{-} \sum_{i}\nabla_{i}\cdot\frac
{1}{\gamma}\mathbf{F}_{\mathbf{r}^{N}}^{\text{tot}\left(  i\right)  }\left[
0,\overline{\Psi}_{t}^{\text{min}}\right]  \overline{\Psi}_{t\mathbf{r}^{N}%
}^\text{{min}}\left[{\overline{\mathbf{J}}_{<t}},\phi\right] \nonumber \\
	& & +\sum_{i}\nabla_{i}\cdot\left(
\frac{\mathbf{J}_{\mathbf{r}_{i}}\left[  0,\overline{\Psi}_{t}^{\text{min}%
}\right]  -\overline{\mathbf{J}}_{t\mathbf{r}_{i}}}{\rho_{t\mathbf{r}_{i}%
}\left[  \overline{\mathbf{J}}\right]  }\;\overline{\Psi}_{t\mathbf{r}^{N}%
}^\text{{min}}\left[{\overline{\mathbf{J}}_{<t}},\phi\right]  \right)  ,\;\;\overline{\Psi}_{t_{0}%
\mathbf{r}^{N}}^\text{{min}}=\Psi_{t_{0}\mathbf{r}^{N}}%
	\label{eq:fieldfree}
\end{eqnarray}
which implies that {$\overline{\Psi}_{t\mathbf{r}^{N}}^{\text{min}}[\overline{\mathbf{J}}_{<t},\phi]
\equiv \overline{\Psi}_{t\mathbf{r}^{N}}^{\text{min}}[\overline{\mathbf{J}}_{<t}]$} is in fact
independent of the external field entirely. Thus, it is, according the usual
terminology of DFT, a universal function\tr{al} (given an initial condition)
satisfying the equation%
\begin{eqnarray}
\frac{\partial}{\partial t}\overline{\Psi}_{t\mathbf{r}^{N}}^{\text{min}%
	}\left[  \overline{\mathbf{J}}\right]  &=&{-} \sum_{i}\nabla_{i}\cdot\frac{1}%
{\gamma}\mathbf{F}_{\mathbf{r}^{N}}^{\text{tot}\left(  i\right)  }\left[
0,\overline{\Psi}_{t}^{\text{min}}\right]  \overline{\Psi}_{t\mathbf{r}^{N}%
}\left[  \overline{\mathbf{J}}\right]  +\nonumber \\
	& &	\sum_{i}\nabla_{i}\cdot\left(
\frac{\mathbf{J}_{\mathbf{r}_{i}}\left[  0,\overline{\Psi}_{t}^{\text{min}%
}\right]  -\overline{\mathbf{J}}_{t\mathbf{r}_{i}}}{\rho_{t\mathbf{r}_{i}%
}\left[  \overline{\mathbf{J}}\right]  }\overline{\Psi}_{t\mathbf{r}^{N}%
}\left[  \overline{\mathbf{J}}\right]  \right) 
	,\;\;\overline{\Psi}%
_{t_{0}\mathbf{r}^{N}}=\Psi_{t_{0}\mathbf{r}^{N}}%
\end{eqnarray}
As already noted, the minimization {with respect to the current for all times} gives%
\begin{equation}
\overline{\mathbf{J}}_{t\mathbf{r}}^{\text{min}}=\mathbf{J}_{\mathbf{r}%
}\left[  \phi,\overline{\Psi}_{t}^{\text{min}}\right]
\end{equation}
which {(using the definitions in Eqs.~(\ref{eq:velocities}) and (\ref{J}))} can be written as%
\begin{equation}
\overline{\mathbf{J}}_{t\mathbf{r}_{i}}^{\text{min}}=\left.-\frac{k_{B}T}{\gamma
}\nabla\rho_{\mathbf{r}_{i}}\left[  \overline{\mathbf{J}}\right]  -\frac
{1}{\gamma}\rho_{\mathbf{r}}\left[  \overline{\mathbf{J}}\right]  \left(
\nabla\phi_{t\mathbf{r}}\right)  +\mathbf{K}_{t\mathbf{r}}\left[
	{\overline{\mathbf{J}}}\right]\right|_{{\overline{\mathbf{J}}=\overline{\mathbf{J}}^{\text{min}}}}
\end{equation}
where the last term is
\begin{equation}
\mathbf{K}_{t\mathbf{r}}\left[  \overline{\mathbf{J}}\right]  =-\frac
{1}{\gamma}\int\sum_{i}\delta\left(  \mathbf{r-r}_{i}\right)  \left(  \left(
\nabla_{i}U\left(  \mathbf{r}^{N}\right)  \right)  \overline{\Psi
}_{t\mathbf{r}^{N}}^{\text{min}}\left[  \overline{\mathbf{J}}\right]  \right)
\;d\mathbf{r}^{N}.
\end{equation}
Although not indicated, everything also depends on the initial condition for
the distribution, $\Psi_{t_{0}\mathbf{r}^{N}}$. Thus, we can summarize the
evolution of the physical density with the following exact equations%
\begin{align}
\frac{\partial}{\partial t}\rho_{t\mathbf{r}}\left[  \overline{\mathbf{J}%
	}^{\text{min}}\right]   &  = {-}\nabla\cdot\overline{\mathbf{J}}_{t\mathbf{r}%
}^{\text{min}}\\
\overline{\mathbf{J}}_{t\mathbf{r}_{i}}^{\text{min}} &  =-\frac{k_{B}T}%
{\gamma}\nabla\rho_{t\mathbf{r}_{i}}\left[  \overline{\mathbf{J}}^{\text{min}%
}\right]  -\frac{1}{\gamma}\rho_{t\mathbf{r}}\left[  \overline{\mathbf{J}%
}^{\text{min}}\right]  \left(  \nabla\phi_{t\mathbf{r}}\right)  +\mathbf{K}%
_{t\mathbf{r}}\left[  \overline{\mathbf{J}}^{\text{min}}\right]  \nonumber
\end{align}
For an ideal gas, $\mathbf{K}_{t\mathbf{r}}=0$ by definition and this is exact
- not as a consequence of an ansatz but directly demonstrated from the
microscopic formalism. In equilibrium, the distribution is constant so
$\overline{\Psi}_{t\mathbf{r}^{N}}^{\text{min}}\left[  \overline{\mathbf{J}%
}\right]  =\Psi_{t_{0}\mathbf{r}^{N}}$ and so
\begin{equation}
\mathbf{K}_{t\mathbf{r}}^{\text{equil}}\left[  \overline{\mathbf{J}}\right]
=-\frac{1}{\gamma}\int\sum_{i}\delta\left(  \mathbf{r-r}_{i}\right)  \left(
\left(  \nabla_{i}U\left(  \mathbf{r}^{N}\right)  \right)  \Psi_{t_{0}%
\mathbf{r}^{N}}^{\text{equil}}\right)  \;d\mathbf{r}^{N}=-\frac{1}{\gamma
}\nabla\frac{\delta F^{\text{(ex)}}\left[  \rho\right]  }{\delta
\rho_{t\mathbf{r}}},
\end{equation}
where $F^{\text{(ex)}}\left[  \rho\right]  $ is the excess Helmholtz free
energy functional of classical DFT. This is again an exact result
(essentially, a consequence of the YBG hierarchy - see, e.g.
Ref.\cite{Archer}). Thus, a natural local-equilibrium assumption would be
to continue to use this form for non-equilibrium systems resulting in
\begin{equation}
\overline{\mathbf{J}}_{t\mathbf{r}_{i}}^{\text{min}}=-\frac{k_{B}T}{\gamma
}\nabla\rho_{t\mathbf{r}_{i}}\left[  \overline{\mathbf{J}}^{\text{min}%
}\right]  -\frac{1}{\gamma}\rho_{t\mathbf{r}}\left[  \overline{\mathbf{J}%
}^{\text{min}}\right]  \left(  \nabla\phi_{t\mathbf{r}}\right)  -\frac
{1}{\gamma}\nabla\left(  \frac{\delta F^{\text{(ex)}}\left[  \rho\right]
}{\delta\rho_{t\mathbf{r}}}\right)  _{\rho_{t\mathbf{r}}\left[  \overline
{\mathbf{J}}^{\text{min}}\right]  }%
\end{equation}
which, when inserted into the continuity equation, gives a closed dynamics for
the density usually referred to as Dynamic Density Functional Theory\ (DDFT).

{
 The investigation of deviations between the true dynamics and the dynamics predicted
 by DDFT (adiabatic dynamics in the sense of a dynamics of quasi--static changes) has been pursued by Schmidt and coworkers
 over the past years, see e.g. \cite{Fortini2014,Stuhlmueller2018}. Even without knowing the explicit form of the equilibrium free energy
 functional the adiabatic dynamics can be determined by simulation \cite{Fortini2014} and thus also the
 nonadiabatic differences to the true dynamics. In terms of the {ansatz} in Eq.~(\ref{eq:Pt}), these
 differences are contained in the excess part $P_t^\text{ex}$ of the functional $P_t[\overline{\mathbf{J}}]$ which is defined as follows:
\begin{eqnarray}
	\label{eq:Ptsplit}
	P_t[\overline{\mathbf{J}}]  =  P_t^\text{(id)}[\overline{\mathbf{J}}] +  P_t^\text{(ex)}[\overline{\mathbf{J}}] 
	 \qquad \text{with} \qquad 
	    P_t^\text{id}[\overline{\mathbf{J}}]  & =& \frac{\gamma}{2}\int    
\frac{\overline{\mathbf{J}}_{t\mathbf{r}}^{2}}{{\rho
}_{t\mathbf{r}}}d\mathbf{r}
\end{eqnarray}
The relation to the functional $K_{t\mathbf{r}}$ defined above is given by
\begin{equation}
  K_{t\mathbf{r}}[\overline{\mathbf{J}}] = \frac{\delta P_t^\text{ex}[\overline{\mathbf{J}}] }{\delta \overline{\mathbf{J}}} 
+\frac{1}{\gamma}\nabla\left(  \frac{\delta F^{\text{(ex)}}\left[  \rho\right]
	}{\delta\rho_{t\mathbf{r}}}\right)_{\rho_{t\mathbf{r}}\left[  \overline{\mathbf{J}}\right]} 
\end{equation}	
Note that we have restricted the functional dependence of the $P_t$ functional to the current only, in line with
our modified formulation of PFT.
}

\section{Conclusions}

We have attempted to clarify the construction of Power Functional\ Theory for
Brownian dynamics. The theory begins with a simple variational principle that
reproduces the exact statistical description of the system based on a
generating function ultimately related to the Rayleigh dissipation function.
The minimization is broken into two steps - first, minimization under
constraints of constant density and current, followed by minimization over
those fields. The first minimization results in a functional of the current
and density fields called the power functional. This is meant to be analogous
to the Helmholtz functional of cDFT and to play a similar role as a starting
point for physically motivated approximations to the dynamics of the full
$N$-body system. Our analysis has identified several problematic points. There
is some ambiguity in the development of SB as to whether the power functional
is actually functional of both density and current fields or only of the
current. A dependence on both seems to lead to unphysical results so that it
should probably be considered as the result of a constrained minimization at
fixed current. Independent of this, the resulting expressions do not have the
structure described by\ SB\ {which in turn has been used in practical applications of the theory
lateron, and thus the status of those is uncertain and the relation to the
underlying power functional is unclear.}

Having identified the source of the problems in the original development, we
have proposed modifications which appear to avoid them. The result is in some
ways very similar to that discussed by\ SB but with one critical
difference:\ in our development it is possible to carry through the first
constrained minimization exactly resulting in the exact power functional. This
turns out to be a rather trivial quadratic function that simply forces the
current to take on its exact expression. The theory therefore seems to provide
no real advantage over, for example, simply introducing approximations
directly into the continuity equation.

More optimistically, we noted that our development does result in an
interesting reformulation of the dynamics whereby a "universal"\ functional of
the current is defined (i.e. one that is independent of the external field)
and that can be used to formulate the dynamics governing the local density.
Dropping unnecessary notation, the resulting theory has the form%
\begin{align}
	\frac{\partial}{\partial t}\rho_{t\mathbf{r}} &  = {-} \nabla\cdot\mathbf{J}%
_{t\mathbf{r}}\\
\mathbf{J}_{t\mathbf{r}} &  =-\frac{k_{B}T}{\gamma}\nabla\rho
_{t\mathbf{r}}-\frac{1}{\gamma}\rho_{t\mathbf{r}}\left(  \nabla
\phi_{t\mathbf{r}}\right)  +\mathbf{K}_{t\mathbf{r}}\left[  \mathbf{J}\right]
\nonumber
\end{align}
with the non-ideal - or excess - contribution to the current given by
\begin{equation}
\mathbf{K}_{t\mathbf{r}}\left[  \mathbf{J}\right]  =-\frac{1}{\gamma}\int
\sum_{i}\delta\left(  \mathbf{r-r}_{i}\right)  \Psi_{t\mathbf{r}^{N}}\left[
\mathbf{J}\right]  \nabla_{i}U\left(  \mathbf{r}^{N}\right)  \;d\mathbf{r}^{N}%
\end{equation}
and finally the "universal" distribution determined by 
{($\Psi_{\mathbf{r}^{N}}^\text{(ini)}$ is the distribution specifying initial conditions)  }
\begin{eqnarray}
\frac{\partial}{\partial t}\Psi_{t\mathbf{r}^{N}}\left[  \mathbf{J}\right]
	&=& {-} \sum_{i}\nabla_{i}\cdot\frac{1}{\gamma}\mathbf{F}_{\mathbf{r}^{N}%
	}^{\text{tot}\left(  i\right)  }\left[  0,{\Psi}\right]  \Psi_{t\mathbf{r}%
^{N}}\left[  \mathbf{J}\right]  +\sum_{i}\nabla_{i}\cdot\left(  \frac
	{\mathbf{J}_{\mathbf{r}_{i}}\left[  0,{\Psi}\right]
-\mathbf{J}_{t\mathbf{r}_{i}}}{\rho_{t\mathbf{r}_{i}}}\Psi_{t\mathbf{r}^{N}%
}\left[  \mathbf{J}\right]  \right) \nonumber \\
	& & \text{with}\;\;\Psi_{t_{0}\mathbf{r}^{N}}%
	=\Psi_{\mathbf{r}^{N}}^\text{(ini)}%
\end{eqnarray}
This is similar to the structure of classical DFT in which one has a universal
functional of the density, the so-called Helmholtz functional $F\left[
\rho\right]  ,$ and from this one constructs simple one-body equations for the
local density given an external field. Here $\mathbf{K}_{t\mathbf{r}}\left[
\mathbf{J}\right]  $ plays the role of the Helmholtz functional and it shares
another characteristic as well:\ its determination depends on solving an
$N$-body problem that is as complicated as simply solving the original
Fokker-Planck equation (or, in the case of cDFT calculating the original
partition function)\ that defined the starting point. Whether there is any
advantage to such a reformulation of the original problem remains to be seen.

\begin{acknowledgments}
The work of JFL\ was supported by the European Space Agency (ESA) and the
Belgian Federal Science Policy Office (BELSPO) in the framework of the PRODEX
	Programme, contract number ESA AO-2004-070. {MO acknowlegdes funding by the
	German Research Foundation (DFG) through grant OE 285\slash 6-1 within the priority program SPP2171.}
\end{acknowledgments}

\section*{Data Availability}

Data sharing is not applicable to this article as no new data were created or
analyzed in this study.

\appendix

\section{Derivation of the form of the generating functional$\label{appA}$}

SB begin by defining (their Eq.(10) )
\begin{equation}
\widehat{R}_{t\mathbf{r}^{N}}\left[  \widetilde{\mathbf{v}}^{N}\right]
=\sum_{i}\left(  \frac{\gamma}{2}\widetilde{\mathbf{v}}_{t\mathbf{r}^{N}%
}^{\left(  i\right)  }-\mathbf{F}_{t\mathbf{r}^{N}}^{\text{tot}\left(
	i\right)  }\right)  \cdot\widetilde{{\mathbf{v}}}_{t\mathbf{r}^{N}}^{\left(
i\right)  }+\overset{\cdot}{\phi}_{t\mathbf{r}^{N}}%
\end{equation}
with (SB Eq.(5))
\begin{equation}
\mathbf{F}_{t\mathbf{r}^{N}}^{\text{tot}\left(  i\right)  }\left[  \phi
_{t},\Psi\right]  =-\nabla_{i}U_{\mathbf{r}^{N}}-\nabla_{i}\phi_{t\mathbf{r}%
^{N}}-k_{B}T\nabla_{i}\ln\Psi_{t\mathbf{r}}%
\end{equation}
and (SB Eq.(17))%
\begin{equation}
\gamma\widetilde{\mathbf{v}}_{t\mathbf{r}^{N}}^{\left(  i\right)  }%
\equiv\gamma\mathbf{v}_{t\mathbf{r}^{N}}^{\left(  i\right)  }\left[  \phi
_{t},\widetilde{\Psi}\right]  =\mathbf{F}_{t\mathbf{r}^{N}}^{\text{tot}\left(
i\right)  }\left[  \phi_{t},\widetilde{\Psi}\right]
\end{equation}
in terms of which the Fokker-Planck equation becomes
\[
\frac{\partial}{\partial t}\Psi_{t\mathbf{r}^{N}}=-\frac{1}{\gamma}\sum
_{i=1}^{N}\mathbf{\nabla}_{i}\cdot\left(  \Psi_{t\mathbf{r}}\mathbf{F}%
_{t\mathbf{r}^{N}}^{\text{tot}\left(  i\right)  }\left[  \phi_{t},\Psi\right]
\right)  .
\]

The functional $\widehat{R}_{t\mathbf{r}^{N}}$ can be written as
\begin{align}
\widehat{R}_{t\mathbf{r}^{N}}\left[  \widetilde{\mathbf{v}}^{N}\right]   &
=\frac{\gamma}{2}\sum_{i}\left(  \widetilde{\mathbf{v}}_{t\mathbf{r}^{N}%
}^{\left(  i\right)  }-\frac{1}{\gamma}\mathbf{F}_{t\mathbf{r}^{N}%
}^{\text{tot}\left(  i\right)  }\right)  ^{2}-\frac{1}{2\gamma}\sum_{i}\left(
\mathbf{F}_{t\mathbf{r}^{N}}^{\text{tot}\left(  i\right)  }\right)
^{2}+\overset{\cdot}{\phi}_{t\mathbf{r}^{N}}\\
&  =\frac{\left(  k_{B}T\right)  ^{2}\gamma}{2}\sum_{i}\left(  \nabla_{i}%
\ln\frac{\widetilde{\Psi}_{\mathbf{r}}}{\Psi_{\mathbf{r}}}\right)  ^{2}%
-\frac{1}{2\gamma}\sum_{i}\left(  \mathbf{F}_{t\mathbf{r}^{N}}^{\text{tot}%
\left(  i\right)  }\right)  ^{2}+\overset{\cdot}{\phi}_{t\mathbf{r}^{N}%
}\nonumber
\end{align}
The SB generating {functional} is (SB Eq.(11))
\begin{align}
	{R} \left[  \widetilde{\Psi};\Psi_{t}\right]   &  =\int
\Psi_{t\mathbf{r}^{N}}\widehat{R}_{t\mathbf{r}^{N}}d\mathbf{r}^{N}%
\label{app1}\\
&  =\frac{\left(  k_{B}T\right)  ^{2}\gamma}{2}\int\sum_{i}\left(  \nabla
_{i}\ln\frac{\widetilde{\Psi}_{\mathbf{r}}}{\Psi_{\mathbf{r}}}\right)
^{2}\Psi_{t\mathbf{r}^{N}}d\mathbf{r}^{N}\nonumber\\
&  -\frac{1}{2\gamma}\int\Psi_{t\mathbf{r}^{N}}\sum_{i}\left(  \mathbf{F}%
_{\mathbf{r}^{N}}^{\text{tot}\left(  i\right)  }\left[  \Psi_{t},\phi
_{t}\right]  \right)  ^{2}d\mathbf{r}^{N}\nonumber\\
&  +\int\Psi_{t\mathbf{r}^{N}}\overset{\cdot}{\phi}_{t\mathbf{r}^{N}%
}d\mathbf{r}^{N}\nonumber
\end{align}
The second term is
\begin{align}
-\frac{1}{2\gamma}\int\Psi_{t}\sum_{i}\left(  \mathbf{F}_{\mathbf{r}^{N}%
}^{\text{tot}\left(  i\right)  }\left[  \Psi_{t},\phi_{t}\right]  \right)
^{2}d\mathbf{r}^{N} &  =\frac{1}{2\gamma}\int\sum_{i}\left(  \nabla_{i}\left(
U_{\mathbf{r}^{N}}+\phi_{t\mathbf{r}^{N}}+k_{B}T\ln\Psi_{t\mathbf{r}^{N}%
}\right)  \right)  \cdot\Psi_{t\mathbf{r}^{N}}\mathbf{F}_{\mathbf{r}^{N}%
}^{\text{tot}\left(  i\right)  }\left[  \Psi_{t},\phi_{t}\right]
d\mathbf{r}^{N}\label{app2}\\
&  =-\frac{1}{2\gamma}\int\left(  U_{\mathbf{r}^{N}}+\phi_{t\mathbf{r}^{N}%
}+k_{B}T\ln\Psi_{t\mathbf{r}^{N}}\right)  \sum_{i}\nabla_{i}\cdot
\Psi_{t\mathbf{r}^{N}}\mathbf{F}_{\mathbf{r}^{N}}^{\text{tot}\left(  i\right)
}\left[  \Psi_{t},\phi_{t}\right]  d\mathbf{r}^{N}\nonumber
\end{align}
where the second line follows from the divergence theorem and surface terms
are ignored. Substituting for the divergence using the Fokker-Planck equation,%
\begin{align}
-\frac{1}{2\gamma}\int\Psi_{t}\sum_{i}\left(  \mathbf{F}_{t\mathbf{r}^{N}%
}^{\text{tot}\left(  i\right)  }\right)  ^{2}d\mathbf{r}^{N} &  =-\frac
{1}{2\gamma}\int\left(  U_{\mathbf{r}^{N}}+\phi_{t\mathbf{r}^{N}}+k_{B}%
T\ln\Psi_{t\mathbf{r}^{N}}\right)  \left(  -\gamma\frac{\partial}{\partial
t}\Psi_{t\mathbf{r}^{N}}\right)  d\mathbf{r}^{N}\\
&  =\frac{1}{2}\frac{\partial}{\partial t}\int\left(  U_{\mathbf{r}^{N}}%
+\phi_{t\mathbf{r}^{N}}+k_{B}T\ln\Psi_{t\mathbf{r}^{N}}\right)  \Psi
_{t\mathbf{r}^{N}}d\mathbf{r}^{N}-\frac{1}{2}\int\left(  \frac{\partial
}{\partial t}\phi_{t\mathbf{r}^{N}}\right)  \Psi_{t\mathbf{r}^{N}}%
d\mathbf{r}^{N}\nonumber\\
&  =\frac{1}{2}\frac{\partial}{\partial t}\Lambda\left[  \phi_{t},\Psi
_{t}\right]  -\frac{1}{2}\int\left(  \frac{\partial}{\partial t}%
\phi_{t\mathbf{r}^{N}}\right)  \Psi_{t\mathbf{r}^{N}}d\mathbf{r}^{N}\nonumber
\end{align}
Putting everything together gives the form used in the main text,
\begin{equation}
	{R} \left[  \widetilde{\Psi};\Psi_{t}\right]  =\frac{\left(
k_{B}T\right)  ^{2}\gamma}{2}\int\sum_{i}\left(  \nabla_{i}\ln\frac
{\widetilde{\Psi}_{t\mathbf{r}^{N}}}{\Psi_{t\mathbf{r}^{N}}}\right)  ^{2}%
\Psi_{t\mathbf{r}^{N}}d\mathbf{r}^{N}+\frac{1}{2}\frac{\partial}{\partial
t}\Lambda\left[  \phi_{t},\Psi_{t}\right]  +\frac{1}{2}\int\Psi_{t\mathbf{r}%
^{N}}\overset{\cdot}{\phi}_{t\mathbf{r}^{N}}d\mathbf{r}^{N}.
\end{equation}

In order to arrive at Eq.(\ref{W}) of the main text, we need to evaluate%
\[
\overline{\mathcal{R}}_{t}\left[  \overline{\rho},\overline{\mathbf{J}}%
;\phi_{t},\overset{\cdot}{\phi}_{t}\right]  =R\left[  \widetilde{\Psi
}^{\text{min}}\left[  \overline{\rho},\overline{\mathbf{J}}\right]  ;\phi
_{t},\overset{\cdot}{\phi}_{t},\overline{\Psi}_{t}\left[  \overline{\rho
},\overline{\mathbf{J}}\right]  \right]
\]
or, using Eq.(\ref{app1}) and dropping all unnecessary functional arguments
for the sake of clarity,
\begin{eqnarray}
\overline{\mathcal{R}}_{t}\left[  \overline{\rho},\overline{\mathbf{J}}%
	;\phi_{t},\overset{\cdot}{\phi}_{t}\right]  &=&\frac{\left(  k_{B}T\right)
^{2}\gamma}{2}\int\sum_{i}\left(  \nabla_{i}\ln\frac{\widetilde{\Psi
}_{t\mathbf{r}}^{\text{min}}}{\overline{\Psi}_{t\mathbf{r}}}\right)
^{2}\overline{\Psi}_{t\mathbf{r}^{N}}d\mathbf{r}^{N} \\
\nonumber	& &-\frac{1}{2\gamma}%
\int\overline{\Psi}_{t\mathbf{r}^{N}}\sum_{i}\left(  \mathbf{F}_{\mathbf{r}%
^{N}}^{\text{tot}\left(  i\right)  }\left[  \overline{\Psi}_{t},\phi
_{t}\right]  \right)  ^{2}d\mathbf{r}^{N}+\int\overline{\Psi}_{t\mathbf{r}%
^{N}}\overset{\cdot}{\phi}_{t\mathbf{r}^{N}}d\mathbf{r}^{N}.
\end{eqnarray}
Now, given the linear dependence of $\mathbf{F}_{\mathbf{r}^{N}}%
^{\text{tot}\left(  i\right)  }$ on the field,%
\begin{equation}
\mathbf{F}_{\mathbf{r}^{N}}^{\text{tot}\left(  i\right)  }\left[  \phi
,\Psi\right]  =\mathbf{F}_{\mathbf{r}^{N}}^{\text{tot}\left(  i\right)
}\left[  0,\Psi\right]  -\nabla_{i}\phi_{\mathbf{r}^{N}}%
\end{equation}
one has that
\begin{align}
-\frac{1}{2\gamma}\int\overline{\Psi}_{\mathbf{r}^{N}}\sum_{i}\left(
\mathbf{F}_{\mathbf{r}^{N}}^{\text{tot}\left(  i\right)  }\left[
\phi,\overline{\Psi}\right]  \right)  ^{2}d\mathbf{r}^{N}  & =-\frac
{1}{2\gamma}\int\overline{\Psi}_{\mathbf{r}^{N}}\sum_{i}\left(  \mathbf{F}%
_{\mathbf{r}^{N}}^{\text{tot}\left(  i\right)  }\left[  0,\overline{\Psi
}\right]  \right)  ^{2}d\mathbf{r}^{N}\\
& +\frac{1}{\gamma}\int\overline{\Psi}_{\mathbf{r}^{N}}\sum_{i}\left(
\mathbf{F}_{\mathbf{r}^{N}}^{\text{tot}\left(  i\right)  }\left[
0,\overline{\Psi}\right]  \cdot\nabla_{i}\phi_{\mathbf{r}_{i}}\right)
d\mathbf{r}^{N}\nonumber\\
& -\frac{1}{2\gamma}\int\overline{\Psi}_{t\mathbf{r}^{N}}\sum_{i}\left(
\nabla_{i}\phi_{\mathbf{r}_{i}}\right)  ^{2}d\mathbf{r}^{N}\nonumber
\end{align}
The second term is
\begin{align}
\frac{1}{\gamma}\int\overline{\Psi}_{\mathbf{r}^{N}}\sum_{i}\left(
\mathbf{F}_{\mathbf{r}^{N}}^{\text{tot}\left(  i\right)  }\left[
0,\overline{\Psi}\right]  \cdot\nabla_{i}\phi_{\mathbf{r}_{i}}\right)
d\mathbf{r}^{N}  & =\int\overline{\Psi}_{\mathbf{r}^{N}}\sum_{i}\left(
\frac{1}{\gamma}\mathbf{F}_{\mathbf{r}^{N}}^{\text{tot}\left(  i\right)
}\left[  0,\overline{\Psi}\right]  \cdot\left\{  \int\left(  \nabla
\phi_{\mathbf{r}}\right)  \delta\left(  \mathbf{r}-\mathbf{r}_{i}\right)
d\mathbf{r}\right\}  \right)  d\mathbf{r}^{N}\\
& =\int\left(  \nabla\phi_{\mathbf{r}}\right)  \cdot\left\{  \int\delta\left(
\mathbf{r}-\mathbf{r}_{i}\right)  \overline{\Psi}_{\mathbf{r}^{N}}\sum
_{i}\left(  \frac{1}{\gamma}\mathbf{F}_{\mathbf{r}^{N}}^{\text{tot}\left(
i\right)  }\left[  0,\overline{\Psi}\right]  \right)  d\mathbf{r}^{N}\right\}
d\mathbf{r}\nonumber\\
& =\int\left(  \nabla\phi_{\mathbf{r}}\right)  \cdot\mathbf{J}\left[
0,\overline{\Psi}\right]  d\mathbf{r}\nonumber
\end{align}
Thus%
\begin{align}
\overline{\mathcal{R}}_{t}\left[  \overline{\rho},\overline{\mathbf{J}}%
;\phi_{t},\overset{\cdot}{\phi}_{t}\right]    & =\frac{\left(  k_{B}T\right)
^{2}\gamma}{2}\int\sum_{i}\left(  \nabla_{i}\ln\frac{\widetilde{\Psi
}_{t\mathbf{r}}^{\text{min}}}{\overline{\Psi}_{t\mathbf{r}}}\right)
^{2}\overline{\Psi}_{t\mathbf{r}^{N}}d\mathbf{r}^{N}-\frac{1}{2\gamma}%
\int\overline{\Psi}_{t\mathbf{r}^{N}}\sum_{i}\left(  \mathbf{F}_{\mathbf{r}%
^{N}}^{\text{tot}\left(  i\right)  }\left[  0,\overline{\Psi}_{t}\right]
\right)  ^{2}d\mathbf{r}^{N}\\
& +\int\left(  \nabla\phi_{t\mathbf{r}}\right)  \cdot\mathbf{J}_{\mathbf{r}%
}\left[  0,\overline{\Psi}_{t}\right]  d\mathbf{r}-\frac{1}{2\gamma}%
\int\left(  \nabla\phi_{t\mathbf{r}}\right)  ^{2}\rho_{\mathbf{r}}\left[
\overline{\Psi}_{t}\right]  d\mathbf{r}^{N}+\int\rho_{\mathbf{r}}\left[
\overline{\Psi}_{t}\right]  \overset{\cdot}{\phi}_{t\mathbf{r}^{N}}%
d\mathbf{r}^{N},\nonumber
\end{align}
{which is Eq.~(\ref{W}) in the main text.}

\section{Minimizing with independent density and current\label{appB}}

To be of use, models must be developed for the functionals to be minimized. SB
propose to write the functional {$W_t^{(SB)}$} as%
\begin{equation}
{W_t^{(SB)}}\left[  \overline{\rho},\overline{\mathbf{J}};\phi_{t},\Psi_{t}\right]
=P^{\left(  \text{ex}\right)  }\left[  \overline{\rho},\overline{\mathbf{J}%
};\phi_{t},\Psi_{t}\right]  +\int\left(  \frac{\overline{\mathbf{J}%
}_{\mathbf{r}^{\prime}}^{2}}{2\gamma\overline{\rho}_{\mathbf{r}^{\prime}}%
}+\overline{\mathbf{J}}_{\mathbf{r}^{\prime}}\cdot\nabla\frac{\delta F\left[
\overline{\rho}\right]  }{\delta\overline{\rho}_{\mathbf{r}^{\prime}}%
}\right)  d\mathbf{r}^{\prime}%
\end{equation}
where $P^{\left(  \text{ex}\right)  }$ is the excess "dissipated power
functional", the quadratic term in $\overline{\mathbf{J}}$ is the "ideal
dissipated power functional" and the third term is the "adiabatic" (i.e. local
equilibrium) contribution which involves the DFT free energy functional
$F\left[  \overline{\rho}_{t}\right]  ${(see Eqs.~(\ref{eq:Pt}) and (\ref{eq:Ptsplit}))}. The Euler-Lagrange equations then become%
\begin{align}
\frac{\delta}{\delta\overline{\rho}_{\mathbf{r}}}P^{\left(  \text{ex}\right)
}\left[  \overline{\rho},\overline{\mathbf{J}};\phi_{t},\Psi_{t}\right]
-\frac{\overline{\mathbf{J}}_{\mathbf{r}}^{2}}{2\gamma\overline{\rho
}_{\mathbf{r}}^{2}}-\int\left(  \nabla^{\prime}\cdot\overline{\mathbf{J}%
}_{\mathbf{r}^{\prime}}\right)  \frac{\delta^{2}F\left[  \overline{\rho
}\right]  }{\delta\overline{\rho}_{\mathbf{r}}\delta\overline{\rho
}_{\mathbf{r}^{\prime}}}d\mathbf{r}^{\prime} &  =0\\
\frac{\delta}{\delta\overline{\mathbf{J}}_{\mathbf{r}}}P^{\left(
\text{ex}\right)  }\left[  \overline{\rho},\overline{\mathbf{J}};\phi_{t}%
,\Psi_{t}\right]  +\frac{\overline{\mathbf{J}}_{\mathbf{r}}}{\gamma
\overline{\rho}_{\mathbf{r}}}+\nabla\frac{\delta F\left[  \overline{\rho
}\right]  }{\delta\overline{\rho}_{\mathbf{r}}}+\nabla\phi_{t\mathbf{r}} &
=0\nonumber
\end{align}
Separating the {equilibrium free energy} functional into its ideal and excess parts
{($F=F^{(\text{id})}+F^{(\text{ex})}$ with 
$F^{(\text{id})}[\rho]=k_BT \int d\mathbf{r} \rho(\mathbf{r})( \ln (\rho(\mathbf{r})\lambda^3) -1) $ where
$\lambda$ is the thermal de--Broglie length)} and rearranging
gives
\begin{align}
\overline{\rho}_{\mathbf{r}}^{2}\frac{\delta}{\delta\overline{\rho
}_{\mathbf{r}}}P^{\left(  \text{ex}\right)  }\left[  \overline{\rho}%
,\overline{\mathbf{J}};\phi_{t},\Psi_{t}\right]  -\frac{1}{2\gamma}%
\overline{\mathbf{J}}_{\mathbf{r}}^{2}-\overline{\rho}_{\mathbf{r}}\left(
\nabla\cdot\overline{\mathbf{J}}_{\mathbf{r}}\right)  -\overline{\rho
}_{\mathbf{r}}^{2}\int\left(  \nabla^{\prime}\cdot\overline{\mathbf{J}%
}_{\mathbf{r}^{\prime}}\right)  \frac{\delta^{2}F^{\left(  \text{ex}\right)
}\left[  \overline{\rho}\right]  }{\delta\overline{\rho}_{\mathbf{r}}%
\delta\overline{\rho}_{\mathbf{r}^{\prime}}}d\mathbf{r}^{\prime} &  =0\\
\overline{\rho}_{\mathbf{r}}\frac{\delta}{\delta\overline{\mathbf{J}%
}_{\mathbf{r}}}P^{\left(  \text{ex}\right)  }\left[  \overline{\rho}%
,\overline{\mathbf{J}};\phi_{t},\Psi_{t}\right]  +\frac{\overline{\mathbf{J}%
}_{\mathbf{r}}}{\gamma}+\nabla\overline{\rho}_{\mathbf{r}}+\overline{\rho
}_{\mathbf{r}}\nabla\frac{\delta F^{\left(  \text{ex}\right)  }\left[
\overline{\rho}\right]  }{\delta\overline{\rho}_{\mathbf{r}}}+\overline{\rho
}_{\mathbf{r}}\nabla\phi_{t\mathbf{r}} &  =0\nonumber
\end{align}
This completes as much of the general framework {as laid out by SB and} needed here and now we
turn to applications to specific systems.

\subsection{Equilibrium}

In equilibrium, there is no time dependence and the current vanishes. We
assume that the contribution from $P^{\left(  \text{ex}\right)  }$ also
vanishes so that the only non-trivial equation is the second (coming from the
variation of the current) which can be written as
\begin{equation}
\overline{\rho}_{\mathbf{r}}\nabla\frac{\delta}{\delta\overline{\rho
}_{\mathbf{r}}}\left(  F\left[  \overline{\rho}\right]  +\int\overline{\rho
}_{\mathbf{r}^{\prime}}\phi_{\mathbf{r}^{\prime}}d\mathbf{r}^{\prime}\right)
=0
\end{equation}
and which has the usual equilibrium DFT solution%
\begin{equation}
F\left[  \overline{\rho}\right]  +\int\overline{\rho}_{\mathbf{r}^{\prime}%
}\phi_{\mathbf{r}^{\prime}}d\mathbf{r}^{\prime}=\text{const.}%
\end{equation}
In this sense, the proposed ansatz for $W$ reproduces equilibrium cDFT.

\subsection{Adiabatic approximation}

By the definition of SB, ``adiabatic" means that one ignores the excess
dissipated power functional {($P^\text{(ex)}=0$)} giving%
\begin{align}
0  &  =\frac{1}{2\gamma}\overline{\mathbf{J}}_{\mathbf{r}}^{2}+\overline{\rho
}_{\mathbf{r}}\left(  \nabla\cdot\overline{\mathbf{J}}_{\mathbf{r}}\right)
+\overline{\rho}_{\mathbf{r}}^{2}\int\left(  \nabla^{\prime}\cdot
\overline{\mathbf{J}}_{\mathbf{r}^{\prime}}\frac{\delta^{2}F^{\left(
\text{ex}\right)  }\left[  \overline{\rho}\right]  }{\delta\overline{\rho
}_{\mathbf{r}}\delta\overline{\rho}_{\mathbf{r}^{\prime}}}\right)
d\mathbf{r}^{\prime}\label{EL-ddft}\\
\frac{1}{\gamma}\overline{\mathbf{J}}_{\mathbf{r}}  &  =-\nabla\overline{\rho
}_{\mathbf{r}}-\overline{\rho}_{\mathbf{r}}\nabla\frac{\delta F^{\left(
\text{ex}\right)  }\left[  \overline{\rho}\right]  }{\delta\overline{\rho
}_{\mathbf{r}}}-\overline{\rho}_{\mathbf{r}}\nabla\phi_{t\mathbf{r}}\nonumber
\end{align}
The second equation gives the current in terms of the density and external
field and if this is inserted into the continuity equation, the resulting
equation of motion can be written as
\begin{equation}
	\frac{\partial}{\partial t}\overline{\rho}_{\mathbf{r}}={+}\nabla\cdot
\overline{\rho}_{\mathbf{r}}\nabla\frac{\delta}{\delta\overline{\rho
}_{\mathbf{r}}}\left(  F\left[  \overline{\rho}\right]  +\int\overline{\rho
}_{\mathbf{r}^{\prime}}\phi_{t\mathbf{r}^{\prime}}d\mathbf{r}^{\prime}\right)
\end{equation}
which looks like the standard DDFT equation of motion. However, as emphasized
above, there is no reason to suppose that the density and current satisfy the
continuity equation and, indeed, as the notation makes clear, $\overline{\rho
}_{\mathbf{r}}$ is a time-independent test-field so that this equation makes
little sense. Rather, the density should be determined from the first of
Eqs.(\ref{EL-ddft}), which is purely local in time. Denoting the solution of
the Euler-Lagrange equation as
\begin{equation}
\overline{\rho}_{\mathbf{r}}^{\ast}\left[  \phi_{t}\right]  , \overline
{\mathbf{J}}_{\mathbf{r}}^{\ast}\left[  \phi_{t}\right]
\end{equation}
since the only time-dependence in Eq.(\ref{EL-ddft}) occurs via the potential,
and substituting into the continuity equation gives
\begin{equation}
\int\frac{\delta\overline{\rho}_{\mathbf{r}}^{\ast}\left[  \phi_{t}\right]
}{\delta\phi_{t\mathbf{r}^{\prime}}}\frac{\partial\phi_{t\mathbf{r}^{\prime}}%
	}{\partial t}d\mathbf{r}^{\prime}={-}\nabla\cdot\overline{\mathbf{J}}%
_{\mathbf{r}}^{\ast}\left[  \phi_{t}\right]
\end{equation}
which can only be satisfied for particular fields. In particular, it is clear
that in the case of an external field that does not depend on time, one could
start with a density which is not the equilibrium density, so that presumably
$\overline{\rho}_{\mathbf{r}}^{\ast}$ will relax to its equilibrium value. In
such a case, the left-hand side vanishes, but $\overline{\mathbf{J}%
}_{\mathbf{r}}^{\ast}$ will be nonzero, and indeed nontrivial, so that the
right hand side does not vanish, thus showing that such a solution cannot be
consistent with the continuity equation. To understand this result better, we
turn to the final special case, the ideal gas.

\subsection{Ideal gas in the adiabatic approximation}

For the ideal gas, the excess part of the free energy functional vanishes
leaving
\begin{align}
\frac{1}{2\gamma}\overline{\mathbf{J}}_{\mathbf{r}}^{2}+\overline{\rho
}_{\mathbf{r}}\left(  \nabla\cdot\overline{\mathbf{J}}_{\mathbf{r}}\right)
&  =0\label{const}\\
\frac{1}{\gamma}\overline{\mathbf{J}}_{\mathbf{r}}  &  =-\nabla\overline{\rho
	}_{\mathbf{r}}-\overline{\rho}_{\mathbf{r}}\nabla{\beta}\phi_{t\mathbf{r}} \qquad {(\beta=1/(k_BT))}.\nonumber
\end{align}
The equation for the current is in fact the exact result which can be derived
directly from the Fokker-Planck equation. It can be used to eliminate the
current from the first equation giving%
\begin{equation}
\overline{\rho}_{\mathbf{r}}\left(  \nabla^{2}\overline{\rho}_{\mathbf{r}%
}\right)  -\frac{1}{2}\left(  \nabla\overline{\rho}_{\mathbf{r}}\right)
^{2}+\overline{\rho}_{\mathbf{r}}^{2}\left(  \nabla^{2}\beta\phi_{t\mathbf{r}%
}-\frac{1}{2}\left(  \nabla\beta\phi_{t\mathbf{r}}\right)  ^{2}\right)  =0
\end{equation}
which is satisfied by the adiabatic solution%
\begin{equation}
\overline{\rho}_{\mathbf{r}}^{\ast}\left[  \phi_{t}\right]  =Ae^{-\beta
\phi_{t\mathbf{r}}}%
\end{equation}
for some constant $A$. However, this gives vanishing current and when
substituted into the continuity equation gives
\begin{equation}
\frac{\partial}{\partial t}\phi_{t\mathbf{r}}=0
\end{equation}
indicating that the minimizers $\overline{\rho}_{\mathbf{r}}^{\ast}$ and
$\overline{\mathbf{J}}_{\mathbf{r}}^{\ast}$ are only consistent with the
continuity equation in the particular case of a time-independent field.

%

\end{document}